\documentclass[prb,twocolumn,showpacs,superscriptaddress,floatfix]{revtex4}
\usepackage{graphicx,amsfonts,amssymb,amsmath,hyperref}

\newif\ifhyper
\hypertrue
\ifhyper
\hypersetup{
 citecolor = {green},
 colorlinks = {true}, 
 urlcolor = {blue} 
}
\fi

\newlength{\ldag}
\def\nbOne{{\mathchoice {\rm 1\mskip-4mu l} {\rm 1\mskip-4mu l} {\rm 
1\mskip-4.5mu l} {\rm
1\mskip-5mu l}}}

\begin{document}

\title{Spontaneous  versus  explicit  replica symmetry breaking in the theory of disordered systems}

\author{D. Mouhanna} 
\email{mouhanna@lptmc.jussieu.fr}
\author{G. Tarjus}
\email{tarjus@lptmc.jussieu.fr}
\affiliation{LPTMC,
CNRS UMR 7600, UPMC, 4 Place Jussieu, 75252 Paris Cedex 05, France}


\begin{abstract}

We investigate the relation between spontaneous and explicit replica symmetry breaking in the theory of disordered systems. On general ground, we prove the equivalence between the replicon operator associated with the stability of the replica symmetric solution in the standard replica scheme and the operator signaling a breakdown of the solution with analytic field dependence in a scheme in which replica symmetry is explicitly broken by applied sources. This opens the possibility to study, via the recently developed functional renormalization group, unresolved questions related to spontaneous replica symmetry breaking and spin-glass behavior in finite-dimensional disordered systems.

\end{abstract}

\pacs{75.10.Nr, 05.10.Cc, 05.70.Fh}

\maketitle

\vspace{0.5cm}

\section{Introduction} 

The presence of quenched disorder in condensed matter systems is known to give rise to a wealth of phenomena, such as new phases and phase transitions, new universality classes in critical behavior, localization, slow and glassy dynamics,  possibly associated with   the proliferation of metastable states. On the theoretical side, quenched disorder requires the introduction of new methodological tools. A main technical difficulty comes from the fact that a disordered system is intrinsically inhomogeneous. As a result, its long-distance properties are described by the whole distribution function of some disorder-dependent free-energy functional. With the exception of rare low-dimensional models for which specific methods can be tailored and of phenomenological approaches, such the droplet picture, that rely on a real-space, hence inhomogeneous, description, most theories rather deal with disorder-averaged quantities, \textit{e.g.} cumulants of a disorder-dependent function or functional. The standard tool to achieve this goal is the ``replica method''.

In its most common use, introduced for the study of spin glasses by Edwards and Anderson \cite{edwards75}, the replica method amounts to computing the average of the logarithm of the disorder-dependent partition function of the system under study by introducing $n$ identical copies of the original system and, once the average over disorder has been performed, by letting the number $n$ of ``replicas'' go to zero with a suitable analytic continuation. The ``replicated'' hamiltonian, or action in field-theoretical language, is invariant under permutations of the replicas, \textit{i.e.} ``replica symmetric''; in the standard implementation, external fields (or sources), when added,  are taken as equal for all replicas so that the full hamiltonian/action including the source terms is also replica symmetric \cite{mezard87}. This replica trick has been mostly used in conjunction with mean-field-like models (fully interacting, infinite-dimensional systems or variables with an infinite number of components) \cite{mezard87,mezard91,young98} and approximations (\textit{e.g.}, the gaussian variational method \cite{giamarchi95,dean97,moskalenko97,tarjus02}). More recently, another variant of the replica method has been considered, in the context of the functional renormalization group (FRG) for random elastic manifold and random field models \cite{fisher85,fisher86,balents93, chauve98,chauve00,chauve01b, ledoussal02b,ledoussal04,ledoussal02,ledoussal03,ledoussal06,ledoussal08b,balents05,ledoussal09b,feldman00,feldman02,tarjus04,tissier06a,tarjus08,tissier08}. In this case, the fundamental variables of the replicated system are coupled to sources that are different for all replicas. The source terms then explicitly break the replica symmetry. This approach provides a convenient procedure to generate the cumulants of the appropriate random functional.

Given the \textit{a priori} different nature of the two above replica schemes, a crucial question is that of their equivalence in the physical limit where all sources are equal. Perturbatively, it is easy to show order by order that  the two procedures provide the same answer. However, it has been realized that perturbation theories fail at describing many of the interesting new phenomena induced by the presence of quenched disorder, as illustrated by the failure of the replica-symmetric solution for the spin-glass phase of the Sherrington-Kirkpatrick (SK) mean-field model \cite{mezard87}  or the breakdown of the ``dimensional reduction'' predictions in random elastic manifolds and in the random field Ising model in low enough dimension \cite{young98}. In the mean-field-like approaches, in which all replicas are coupled to the same source, the way out of the naive, perturbative solution comes from a spontaneous breaking of the replica symmetry in the continuation process of taking the number of replicas to zero, as in the Parisi solution of the SK spin glass model \cite{mezard87}. In the FRG description of random manifold and random field systems, in which replica symmetry is explicitly broken from the beginning by the sources, evading dimensional reduction proceeds through the appearance of a nonanalyticity in the field dependence of the running renormalized cumulants of the random free-energy functional as the RG flow approaches a zero-temperature fixed point  \cite{fisher85,fisher86,balents93, chauve98,chauve00,chauve01b,ledoussal02b,ledoussal04,ledoussal02,ledoussal03,ledoussal06,ledoussal08b,balents05,ledoussal09b,feldman00,feldman02,tarjus04,tissier06a,tarjus08,tissier08}.  This nonanalyticity shows up in the limit where the difference between the sources acting on the replicas go to zero. Obviously, the perturbative equivalence between the two replica approaches, with and without explicit breaking of the permutation symmetry, becomes questionable when either one of the approaches encounter a singularity.

The purpose of the present article is then to investigate the relation between spontaneous and explicit replica symmetry breaking in the theoretical description of disordered systems. This topic has already been studied in a series of articles on random elastic manifolds, starting with the work of Balents, Bouchaud and M\'ezard \cite{balents96}. In particular, a thorough analysis has been carried out by Le Doussal,  Wiese {\it et  al}\,  \cite{ledoussal02,ledoussal03,ledoussal08b}   in the large $N$ (infinite codimension) limit of the model. In this specific case, the authors have been able to unravel the full connection between the spontaneously generated replica symmetry broken solution previously obtained by M\'ezard and Parisi \cite{mezard91} and the nonanalytic field dependence of the cumulants obtained by combining FRG and explicit replica symmetry breaking (RSB). The question we address in the following is whether a similar connection can be established for a more general class of disordered systems and for more general, non mean-field, conditions. Such a connection would be especially valuable in the case of the random field $O(N)$ model for which the existence of a spontaneous RSB phenomenon has been suggested \cite{mezard92,mezard94,dedominicis95,pastor02} whereas recent FRG studies have shown that a cusp singularity in the field dependence of the cumulants of the renormalized random field appear below a critical dimension \cite{tarjus04,tissier06a,tissier06b,tarjus08,tissier08}. It would be interesting too for spin glasses. Indeed, the nature of the low-temperature spin-glass phase as well as the existence of a transition in the presence of an applied magnetic field are still controversial matters. Combining the FRG approach and the explicit RSB scheme could therefore shed light on the presence or absence of spontaneous RSB in the system.

When found in theoretical descriptions of disordered models, spontaneous RSB in its most common form involves a ``matrix breaking'', \textit{i.e.} involves a 2-point, 2-replica function, as in the Parisi solution of the SK model \cite{mezard87}. As developed in the theory of spin glasses and discussed more recently for the random field $O(N)$ model \cite{mezard92,mezard94,dedominicis95,pastor02},  spontaneous RSB is signaled by an instability of the replica-symmetric solution which is characterized by the appearance of a zero eigenvalue in an appropriate stability operator. The natural framework to investigate such phenomena involving 2-point, 2-replica functions is the so-called 2-particle irreducible (2PI) formalism \cite{luttinger60,baym62,cornwall73}: by introducing  sources that couple not only to the fundamental fields but also to composite bilinear fields and then performing a double Legendre transform, one obtains a generating functional that has  for argument both  fields (magnetizations) and  2-point correlation functions.   When coupled to a replica approach, this formalism allows one to write the above mentioned stability operator as the jacobian of the double Legendre transform \cite{dedominicis95, dedominicis06}. (On the other hand, the FRG approach to  disordered systems is commonly expressed    within the  1-particle irreducible (1PI)  formalism in  which  sources are linearly coupled to the fundamental fields;  the 1PI formalism can be recovered  from the 2PI  one,  which makes  the latter a suitable starting point for studying the relation between  spontaneous and explicit RSB.)  In this work, we show  that a general relation can be derived between the instability commonly associated with a (continuous) spontaneous RSB and the appearance of a nonanalytic field dependence in the context of the explicit RSB approach.  This relation suggests  a new way to tackle putative spontaneous  RSB in disordered systems, such as  spin glasses,  by means of a general FRG approach.    In addition,  we illustrate this  relation in the context of  the random field $O(N)$ model,  in which  we  establish explicit expressions  at leading and next-to-leading orders of the $1/N$ expansion.

The article is organized as follows. In section II, we introduce the 2PI formalism in the context of the replica method for disordered systems. We next compare in section III the replica symmetric and explicit RSB formal solutions for the 2-point, 2-replica correlation functions. Section IV contains the main result of this study, a relation between the stability operators that characterize spontaneous RSB on the one hand and breaking of analyticity in the explicit RSB formalism on the other hand. Section V further expands on the consequences of this relation and provides some concluding remarks.   In appendix A we   illustrate   the  relation between stability operators  in the context of the random field $O(N)$ model in the large $N$ limit and in appendix B  we consider the case of local self-energies. 

\section{2PI formalism with replicas}

The  2PI formalism \cite{luttinger60,baym62,cornwall73}   involves a functional, an ``effective action'', that depends on two kinds of order parameters.  The first one is the usual  order parameter  $\phi$ --   the magnetization -- which is relevant to characterize   the occurence of a paramagnetic-ferromagnetic transition.  The second type of order parameter consists of 2-point correlation functions. To implement this formalism, one has to introduce external sources that not only linearly couple to the fundamental variables (as a magnetic field) but also couple to composite fields formed by the product of two fundamental variables.

We consider a system originally defined on a lattice and described, at some coarse-grained  scale $\Lambda$ ,  by a  continuous field $\chi(x)$.  This field represents a  spin variable for random magnets,  a displacement  field  parametrizing  a manifold  in the case of  an elastic manifold pinned by disorder, etc.  We restrict ourselves to a scalar field and for simplicity we write down the equations for a one-component field; extension to many components is straightforward. To this system  is associated    an effective Hamiltonian or action  $S[\chi;h] $ where  $h$ is a symbolic notation  to characterize the presence of quenched disorder (dilution, random couplings, random magnetic field or anisotropy, random potential, etc).  The partition function of the system for a given sample with disorder $h$ is given by 
\begin{equation}
\begin{array}{ll}
\displaystyle Z_h[ J, K]=  \displaystyle \int {\cal D} \chi& \exp\bigg(-S[\chi;h] +  \displaystyle \int_x   J(x) \chi(x)\\
\\&\displaystyle +  \frac{1}{2}   \int_x \int_y  \chi(x) K(x,y) \chi(y)\bigg),
\label{ZjK}
\end{array}
\end{equation}
where  the scalar field  $\chi$ is coupled  linearly to a source  $J$  and  quadratically to a source $K$, and where $\int_x$ denotes $\int d^d x$. A random, \textit{i.e.} disorder-dependent, free-energy functional $W_h[J,K]$ can be defined as the logarithm of $Z_h[ J, K]$.

Within the replica formalism, one introduces $n$ replicas  of the original system. We consider the general case in which the sources are different for all replicas and pairs of replicas. After performing the average over the disorder, one obtains
\begin{equation}
\begin{array}{ll}
\displaystyle Z_{rep}[ \{J_{a}\}, \{K_{ab}\}]&= \displaystyle \int \prod_{a=1}^n  {\cal D}  \chi_{a}\  {\hbox{exp}} \bigg( - S_{rep}[\{\chi_{a}\}]  \\
\\&\displaystyle + \sum_{a=1}^n \int_x  J_{a}(x)  \chi_{a}(x)  \\ \\& \displaystyle + \frac{1}{2} \sum_{a,b}^n \int_x \int_y \chi_{a}(x) K_{ab}(x,y) \chi_{b}(y)\bigg)
\label{ZjKrep}
\end{array}
\end{equation}
where Latin letters denote replica indices and $S_{rep}[\{\chi_{a}\}]$  can be generically written as
\begin{equation}
\begin{array}{ll}
\displaystyle S_{rep}[\{\chi_{a}\}] &= \displaystyle \sum_{a=1}^n \int_x\, S_1[\chi_{a}] \\
\\
&\displaystyle - \frac{1}{2}  \sum_{a,b}^n \int_x \int_y \,  S_2[\chi_{a},  \chi_{b}]+\dots
\label{replicated_action}
\end{array}
\end{equation}
with $S_1[\chi_{a}] = \overline{S[\chi_{a};h]}$, $S_2[\chi_{a},  \chi_{b}] = \overline{S[\chi_{b};h]S[\chi_{a};h]}-\overline{S[\chi_{a};h]}\;\overline{S[\chi_{b};h]}$, etc; an overbar as usual denotes an average over the disorder $h$. Higher-order terms in the number of sums over replicas occur when nongaussian distributions of the disorder are considered.  The specific form of $S_1$, $S_2$, etc, of course varies with the kind of model -- random field, random anistropy, random temperature, random manifold, spin glass, etc --  considered, but at this point, we keep the formalism in its full generality.

From the partition function of the ``replicated'' system in Eq.~(\ref{ZjKrep}) one defines the free-energy functional $W_{rep}[ \{J_{a}\}, \{K_{ab}\}]=\ln  Z_{rep}[ \{J_{a}\}, \{K_{ab}\}]$ 
whose first derivatives generate the usual ferromagnetic order parameter,
\begin{equation}
\displaystyle {\delta W_{rep}[ \{J_{e}\}, \{K_{e f}\}]\over\delta J_{a}(x)}=\langle\chi_{a}(x)\rangle =\phi_{a}(x), 
\label{phi2}
\end{equation}
where $\langle\, \rangle$ denotes  a thermal average with the replicated action, and the 2-point, 2-replica correlation functions,
\begin{equation}
\begin{array}{ll}
\displaystyle{\delta W_{rep}[ \{J_{e}\}, \{K_{e f}\}]\over\delta  K_{ab}(x,y)}&=\displaystyle{1\over 2} \langle\chi_{a}(x)\chi_{b}(y)\rangle \\
\\
& =\displaystyle {1\over 2} \big(G_{ab}(x,y)+\phi_{a}(x)\phi_{b}(y)\big),
\label{G}
\end{array}
\end{equation}
where  $G_{ab}(x,y)=\langle\chi_{a}(x)\chi_{b}(y)\rangle-\langle\chi_{a}(x)\rangle\langle\chi_{b}(y)\rangle$ is the connected 2-point Green function (or propagator). The latter is also obtained through two derivatives of $W_{rep}$ with respect to $J_{a}(x)$ and $J_{b}(y)$. Higher-order derivatives with respect to the sources generate higher-order connected Green functions.

The double Legendre transform with respect to the two sources $J_{a}$ and $K_{ab}$ defines an effective action $\Gamma_{2PI}[\{\phi_{a}\},\{G_{ab}\}]$:
\begin{equation}
\begin{array}{ll}
&\displaystyle  \Gamma_{2PI}[\{\phi_{a}\},\{G_{ab}\}] =\\
\\& \displaystyle -W_{rep}[\{J_{a}\},  \{K_{ab}\}] + \sum_{a=1}^n  \int_x  J_{a} (x) \phi_{a}(x) \\
\\& \displaystyle  + \frac{1}{2} \sum_{a,b=1}^n   \int_x \int_y  K_{ab}(x,y) \big(G_{ab}(x,y) + \phi_{a}(x) \phi_{b}(x) \big),
\label{gamma2PI}
\end{array}
\end{equation}
where we have used Eqs.~(\ref{phi2}) and (\ref{G}). Differentiation with respect to the field $\phi_{a}$ and to the 2-point, 2-replica correlation function  $G_{ab}$ provides
\begin{eqnarray}
 \hspace{-0cm} \displaystyle   {\delta \Gamma_{2PI}[\{\phi_{e}\},\{G_{ef}\}]\over \delta \phi_{a}(x)}&=& \nonumber \\
\nonumber\\
\hspace{-1cm}   J_{a}(x) &+ & \displaystyle  \sum_{b} \int_y \  K_{ab}(x,y) \phi_{b}(y)\\
\label{statio1}
\nonumber \\
\nonumber \\
& & \hspace{-3.7cm}  \displaystyle  {\delta \Gamma_{2PI}[ \{\phi_{e}\},\{G_{ef}\}]\over \delta G_{ab}(x,y)}= \displaystyle   {1\over 2} \ K_{ab}(x,y)\  .
\label{statio2}
\end{eqnarray}

From now on we consider the case where  all sources  $\{K_{ab}\}$  are set equal to zero.  Eq.~(\ref{statio2}) provides the ``equation of motion'' or Schwinger-Dyson equation for the  fields $\{G_{ab}\}$ which  can be formally solved  in terms of  the fields $\{\phi_{e}\}$: $G_{ab}=G_{ab}^*[\{\phi_{e}\}] $, with
\begin{equation}
\displaystyle {\delta \Gamma_{2PI}[\{\phi_{e}\},\{G_{ef}\}]\over \delta G_{ab}(x,y)}\bigg\vert_{G^*}=  0\ .
\label{saddlepoint}
\end{equation}\\
Substituting this result into the expression of   $\Gamma_{2PI}[\{\phi_{a}\},\{G_{ab}\}]$ leads to the  usual, one-particle irreducible (1PI), effective action:

 \begin{equation}
\displaystyle \Gamma_{2PI}[ \{\phi_{a}\},\{G^*_{ab}[\{\phi_{a}\}]\}]=\Gamma_{1PI}[\{\phi_{a}\}]\ . 
\label{1PI}
\end{equation}
\\
This set of  equations  provides  the starting point  of our approach.   Indeed,  Eq.~(\ref{saddlepoint}), when evaluated for identical sources (or in a more restrictive setting for sources and/or fields $\phi_{a}$ all equal to zero), provides a self-consistent equation for $G_{ab}$  which in limiting (mean-field) cases or with approximations (see below) can be used to study the disordered model at hand and the occurence of spontaneous RSB. On the other hand, when keeping the sources $J_{a}$ acting on the various replicas all different, the reduction to the 1PI effective action in Eq.~(\ref{1PI}) suggest the possibility of including the FRG studies in the same framework.

The 2PI effective action is generally written, up to an additive constant, as \cite{cornwall73}
\begin{equation}
\begin{array}{ll}
\Gamma_{2PI}[\{\phi_{a}\},\{G_{ab}\}]&=S_{rep}[\{\phi_{a}\}]+\displaystyle {1\over 2} \hbox{Tr} \ln G^{-1}+\\
\\ & \hspace{-1.5cm}\displaystyle +  {1\over 2}\hbox{Tr}\,  G\,G_0^{-1}(\{\phi_{a}\}) +\  \Gamma_2[\{\phi_{a}\},\{G_{ab}\}],\
 \end{array}
\label{Gamma2PI}
\end{equation}
where the trace involves a sum over replica indices and an integration over space; $G_0^{-1}$ is the classical (bare) inverse propagator: 
\begin{equation}
\displaystyle \left[ G_{0}^{-1}(x,y;\{\phi_{e}\})\right]_{ab}={\delta^2S_{rep}[\{\phi_{e}\}]\over \delta \phi_a(x)\delta \phi_b(y)}
\end{equation}
and   $\Gamma_2[\{\phi_{a}\},\{G_{ab}\}]$ includes all  2PI  components of  the effective action.  To see this,  one derives  Eq.~(\ref{Gamma2PI})   with respect to the $G_{ab}$'s (again at vanishing sources $K_{ab}$). One finds
\begin{equation}
\begin{array}{ll}
\left[ G^{*-1}(\{\phi_{e}\})\right]_{ab}=&\left[G_{0}^{-1}(\{\phi_{e}\})\right] _{ab} \\ \\&- \Sigma_{ab}\left[ \{\phi_{e}\}; \{G^{*}_{cd}(\{\phi_{e}\})\}\right] 
 \end{array}
\label{GG}
\end{equation}
where  we have defined 
\begin{equation}
\Sigma_{ab}\left[ x,y;\{\phi_{e}\},\{G_{cd}\}\right]  \equiv - 2 {\delta \Gamma_2 [ \{\phi_{e}\},\{G_{cd}\}]\over \delta G_{ab}(x,y)},
\label{selfsigma}
\end{equation}
which is nothing but the self-energy for the replicated system.  As well known, the self-energy $\Sigma$ is made up  of  1PI diagrams, and a functional derivative with respect to $G$  corresponds to opening one propagator line. This  implies that  $\Gamma_2$ is made up of 2PI vacuum diagrams only, with dressed propagator $G$. 

The 2PI functional is of course not known exactly except for very specific conditions. Approximations are thus required for practical computations on disordered systems. In many approximations used in the literature, the self-energy is taken as purely local and all momentum dependences are contained in the bare inverse propagator $G_0^{-1}$. This is verified in mean-field models \cite{mezard87, mezard91} and is the case for (i) the Gaussian variational method which amounts to replacing the original Hamiltonian by a variational Gaussian one \cite{giamarchi95,dean97,moskalenko97,tarjus02}  (and is exact in the limit of large number $N$ of spin components \cite{mezard91})  and (ii) for the ``locator'' approximation \cite{feigelman79,bray79,lopatin02,muller04} and related dynamical mean-field theory (DMFT) approaches  \cite{georges96}, in which the local part of the self-energy is calculated through an effective single site model. Other types of approximations involve expansions either in loops or in powers of $1/N$ in the large $N$ limit. We provide examples in Appendices A and B.

\section{Replica symmetric and explicit RSB solutions}

Consider first the situation in which all sources $J_{a}$ are equal, $J_{a}=J$ (recall that all sources $K_{ab}$ are taken to zero). The replicated action including the source terms is now invariant under permutation of the replicas. When no spontaneous breaking of replica symmetry occurs, the solution of the (Schwinger-Dyson) equation of motion, Eq.~(\ref {GG}), is replica symmetric (RS), namely $\phi_{a}=\phi$,  $\forall a$,  $G_{ab}[\phi]=G_C[\phi] \delta_{ab} + G_D[\phi]$, where $G_C$ and $G_D$ as usual denote the (disorder) connected and disconnected pair correlation functions (from now on we drop the star that characterizes the solution). The algebra of RS matrices is rather simple (see the above form of $G_{ab}$) and by further restricting our study to a uniform source, Eq.~(\ref {GG}) can be rewritten as
\begin{equation}
G_C^{-1}(q;\phi )=G_{0C}^{-1}(q;\phi)-\Sigma_{C}\left[ q;\phi;G_C(\phi),G_D(\phi)\right],
\label{G_C}
\end{equation}
\begin{equation}
G_D^{-1}(q;\phi )=G_{0D}^{-1}(q;\phi)-\Sigma_{D}\left[ q;\phi;G_C(\phi),G_D(\phi)\right],
\label{G_D}
\end{equation}
where we have also decomposed the RS self-energy according to
\begin{equation}
\begin{split}
\Sigma_{ab}\left[ q;\phi;G_C (\phi),G_D(\phi)\right] =\ &\Sigma_C\left[ q;\phi;G_C(\phi),G_D(\phi)\right] \, \delta_{ab}\\ 
\\
& +\Sigma_D\left[ q;\phi;G_C(\phi),G_D(\phi)\right] 
\end{split}
\label{selfsigma_C,D}
\end{equation}
and have kept square brackets to indicate that the self-energies are still functionals of the correlation functions $G_C$, $G_D$. The inverse connected and disconnected  functions (\textit{i.e.}, the connected and disconnected 2-point proper vertices) are given,  in the limit where the number of replicas goes to zero, Êby
\begin{equation}
G_C^{-1}(q;\phi )=G_C(q;\phi)^{-1},
\label{inverseG_C}
\end{equation}
\begin{equation}
G_D^{-1}(q;\phi )=- G_C(q;\phi)^{-2} G_D(q;\phi).
\label{inverseG_D}
\end{equation}

The solution of the Schwinger-Dyson equation in the presence of an explicit RSB due to the sources $J_{a}$ is \textit{a priori} much harder to handle, even at a formal level. However, a workable scheme is provided by using systematic expansions in increasing number of unrestricted sums over replicas (or ``free replica sums'') \cite{balents93, ledoussal02, ledoussal03,ledoussal04,ledoussal04b,tarjus04,tissier06a,tarjus08,tissier08}. For instance, any function of the replica fields $\phi_{a}$, say a function $F(\{\phi_{a}\})$, can be expanded as
\begin{equation}
F(\{\phi_{a}\}) = \sum_{p\geq 1}\frac{1}{p!} \sum_{a_1,...,a_p}F_p(\phi_{a_1},...,\phi_{a_p})
\label{expansion_function}
\end{equation}
where the $F_p$'s are continuous and symmetric functions of their arguments, whereas matrices, say a matrix $A_{ab}(\{\phi_{e}\})$ can be written as 
\begin{equation}
A_{ab}(\{\phi_{e}\})=\widehat{A}_{a}(\{\phi_{e}\})\, \delta_{a b} + \widetilde{A}_{a b}(\{\phi_{e}\}),
\label{expansion_matrix}
\end{equation}
where $\widetilde{A}_{ab}$ does not contain any Kronecker symbol and
\begin{equation}
\begin{split}
\widehat{A}_{a}(\{\phi_{e}\}) = &\widehat{A}^{[0]}(\phi_{a}) \\&+ \sum_{p\geq 1}\frac{1}{p!} \sum_{e_1,...,e_p}\widehat{A}^{[p]}(\phi_{a}|\phi_{e_1},...,\phi_{e_p}),
\label{expansion_hatA}
\end{split}
\end{equation}
\begin{equation}
\begin{split}
\widetilde{A}_{ab}(\{\phi_{e}\}) =& \widetilde{A}^{[0]}(\phi_{a},\phi_{b}) \\&+ \sum_{p\geq 1}\frac{1}{p!} \sum_{e_1,...,e_p}\widetilde{A}^{[p]}(\phi_{a},\phi_{b}|\phi_{e_1},...,\phi_{e_p}).
\end{split}
\label{expansion_tildeA}
\end{equation}
Higher-order tensors can be expanded as well \cite{ledoussal04b}. Terms appearing in the various equations can then be expanded in number of free replica sums and, after properly collecting and symmetrizing, one can proceed to an order-by-order identification.

Applied to Eq.~(\ref{GG}) and again to uniform sources $J_{a}$, the procedure leads at zeroth-order to the following expressions:
\begin{equation}
\begin{split}
\widehat{G}^{[0]-1}(q;\phi_{a})=\widehat{G}_0^{[0]-1}(q;\phi_{a})- \widehat{\Sigma}^{[0]}(q;\phi_{a}),
\label{widehatG0}
\end{split}
\end{equation}
\begin{equation}
\begin{split}
\widetilde{G}^{[0]-1}(q;\phi_{a},\phi_{b})=\widetilde{G}_0^{[0]-1}(q;\phi_{a},\phi_{b}) -  \widetilde{\Sigma}^{[0]}(q;\phi_{a},\phi_{b}),
\end{split}
\label{widetildeG0}
\end{equation}
where
\begin{equation}
\widehat{\Sigma}^{[0]}(q;\phi_{a}) \equiv \widehat{\Sigma}^{[0]}\bigg[ q;\phi_{a};\widehat{G}^{[0]}(\phi_{a}),\widetilde{G}^{[0]}(\phi_{a},\phi_{a})\bigg],
\label{widehatSigma0}
\end{equation}
\begin{equation}
\begin{split}
\widetilde{\Sigma}^{[0]}(q;\phi_{a},\phi_{b}) \equiv \; &\widetilde{\Sigma}^{[0]}\bigg[ q;\phi_{a},\phi_{b};\widehat{G}^{[0]}(\phi_{a}),\widehat{G}^{[0]}(\phi_{b}),\\&\widetilde{G}^{[0]}(\phi_{a},\phi_{a}), \widetilde{G}^{[0]}(\phi_{b},\phi_{b}),\widetilde{G}^{[0]}(\phi_{a},\phi_{b})\bigg],
\end{split}
\label{widetildeSigma0}
\end{equation}
and the inverse functions $\widehat{G}^{[0]-1},\widetilde{G}^{[0]-1}$ are given by
\begin{equation}
\widehat{G}^{[0]-1}(q;\phi_{a})=\widehat{G}^{[0]}(q;\phi_{a})^{-1},
\label{inverse_widehatG0}
\end{equation}
\begin{equation}
\begin{split}
\widetilde{G}^{[0]-1}&(q;\phi_{a},\phi_{b})=\\&- \widehat{G}^{[0]}(q;\phi_{a})^{-1} \widetilde{G}^{[0]}(q;\phi_{a},\phi_{b})\widehat{G}^{[0]}(q;\phi_{b})^{-1}.
\end{split}
\label{inverse_widetildeG0}
\end{equation}
Note that $\widehat\Sigma^{[0]}$ and $\widetilde\Sigma^{[0]}$ are still functionals of the 2-point functions $\widehat{G}^{[0]}$ and $\widetilde{G}^{[0]}$.  However, an important observation is that the replica fields  $\phi_a$, $\phi_b$  that are arguments of the self-energies in   the left-hand side of Eqs.~(\ref{widehatSigma0},\ref{widetildeSigma0}) are the same as those  that are arguments of  the correlation functions entering in the functionals of  the right-hand side of Eqs.~(\ref{widehatSigma0},\ref{widetildeSigma0}). This is a direct consequence of the expansion in free replica sums. In addition, the functions depending on 2-replica fields, $\widetilde{G}^{[0]}(q;\phi_{a},\phi_{b})$ and the associate proper vertex and self-energy, are invariant under   the exchange of the arguments and are therefore even in  $\phi_{a}-\phi_{b}$.

Suppose now that all replica fields \{$\phi_a$\} are equal, $\phi_a=\phi, \;\forall a$. Provided that the dependence of the functions on the replica field arguments is analytic when the arguments become equal, any matrix $A_{ab}(\{\phi_{e}\})$ is RS. As a consequence, it can as well be written as $A_C(\phi) \delta_{ab} + A_D(\phi)$ and, in the limit where the number of replicas goes to zero, one finds that 
\begin{equation}
\widehat{A}_{a}(\{\phi_{e}=\phi\})= \widehat{A}^{[0]}(\phi) =A_C(\phi),
\end{equation}
\begin{equation}
\widetilde{A}_{ab}(\{\phi_{e}=\phi\}) = \widetilde{A}^{[0]}(\phi,\phi)=A_D(\phi).
\end{equation}
In particular, $\widehat{G}^{[0]}(q;\phi)=G_C(q;\phi)$, $\widetilde{G}^{[0]}(q;\phi,\phi)=G_D(q;\phi)$, and Eqs.~(\ref{widehatG0}-\ref{inverse_widetildeG0}) coincide with Eqs.~(\ref{G_C}-\ref{inverseG_D}). This property, as we shall delve on more below, relies on the assumption of analytic field dependence in the limit where all fields are equal. If not the case, Eqs.~(\ref{widehatG0}-\ref{widetildeSigma0}) are {\it a priori}  more general than the RS equations, Eqs.~(\ref{G_C}-\ref{selfsigma_C,D}).

\section{Stability considerations}

We now establish  the main result of this article, \textit{i.e.} the relation between the existence of an instability of the RS solution for the 2-point, 2-replica functions and the appearance of a nonanalyticity 
in the field dependence of these functions in the explicit RSB approach. The 2PI formalism allows one to investigate the stability of the solutions to the Schwinger-Dyson equations, either in the RS scheme [Eqs.~(\ref{G_C}-\ref{selfsigma_C,D})] or in the explicit RSB scheme  [Eqs.~(\ref{widehatG0}-\ref{widetildeSigma0})].

\subsection{Stability of the RS solution}

We start by looking at the stability of the RS solution when $n\rightarrow 0$. This involves considering the eigenvalues of the matrix formed by all the second functional derivatives of $\Gamma_{2PI}$ with respect to the $\phi_a$'s and $G_{ab}$'s. As is well documented since the early work of de Almeida and Thouless on the Sherrington-Kirkpatrick mean-field spin glass model \cite{almeida78}, an instability with respect to a continuous breaking of replica symmetry is signaled by the fact that the stability matrix in the so-called ``replicon sector'' is no longer strictly positive definite. Restricting ourselves to the study of this replicon component, we only need the matrix operator
\begin{equation}
\begin{split}
\displaystyle \Lambda_{ab,cd}(q,q')=&\,2 \displaystyle \int_{q''} \int_{q'''}\delta(q'+q''') \\ \\& \times \frac{\delta^2  \Gamma_{2PI}[\{\phi_{e}\}, \{G_{ef}\}]}{ \delta G_{ab}(q,q'')\delta G_{cd}(q',q''')}\bigg\vert_{RS} \,,         
\end{split}
\label{matrixM}
\end{equation}
where the subscript $RS$ indicates that the derivative is evaluated at the RS solution of Eqs.~(\ref{G_C}-\ref{selfsigma_C,D}),  $\int_{q}$ denotes $\int d^dq/(2\pi)^d$, and the replica indices are such that $a\neq b$ and $c\neq d$; the factor of 2 is introduced for further convenience (in the derivation in the right-hand side of Eq.~(\ref{matrixM}), $G_{ab}$ and $G_{ba}$ are first formally  considered as different). With uniform souces $J_a$ and vanishing $K_{ab}$'s, the system is translationally  invariant which brings a global factor $\delta(q+q'+q''+q''')$ in the second derivative of $\Gamma_{2PI}$; the delta factor in the right-hand side could therefore be replaced by $\delta(q+q'')$ as well.  It is  convenient to consider a symmetrized form of $\Lambda_{ab,cd}$ which, due to replica symmetry, has only $3$ different components,
\begin{equation}
\begin{array}{ll}
\Lambda_1(q,q')=\textrm{sym}\{\Lambda_{ab,ab}(q,q')\}\\
\\
\Lambda_2(q,q')=\textrm{sym}\{\Lambda_{ab,ac}(q,q')\}\\
\\
\Lambda_3(q,q')=\textrm{sym}\{\Lambda_{ab,cd}(q,q')\}\\
\end{array}
\label{Lamdda_i's}
\end{equation}
where $\textrm{sym}\{\Lambda_{ab,ab}\}$ indicates $(1/4)(\Lambda_{ab,ab}+\Lambda_{ba,ab}+\Lambda_{ab,ba}+\Lambda_{ba,ba})$, etc; $\Lambda_1,\Lambda_2$ and $\Lambda_3$ respectively involve 2, 3 and 4 distinct replica indices. The replicon eigenvalue of the matrix is then given by \cite{almeida78}
\begin{equation}
\Lambda_R(q,q')=\Lambda_1(q,q')-2\Lambda_2(q,q')+\Lambda_3(q,q').
\label{replicon}
\end{equation}
Note that in the rather generic case considered here, the replicon is an operator that is momentum dependent. It is however not the most general form as we have somewhat restricted the momentum dependence in Eq.~(\ref{matrixM}).   This is nonetheless sufficient when the system of interest is statistically uniform (which is the case of the replicated system described by Eq.~(\ref{ZjKrep})) with uniform and translationally invariant applied sources \cite{dedominicis95}.

The replicon operator is definite positive in the paramagnetic phase that is generically present at high enough temperature and disorder strength. The vanishing of its smallest eigenvalue, if it indeed occurs, defines a generalization of the de Almeida-Thouless line of instability found in the  SK spin-glass model \cite{almeida78}. In the latter case, it characterizes the instability of the RS solution toward spontaneous RSB. Such an interpretation also applies to the random elastic manifold model in the large $N$ limit (\textit{i.e.}, infinite codimension for the manifold) \cite{ledoussal02,ledoussal03,ledoussal04b}   and more generally to the systems studied via either the gaussian variational method \cite{giamarchi95, tarjus02} or the locator approximation \cite{feigelman79,bray79,lopatin02,muller04}. In the above cases though, the mean-field-like character of the problem or of the approximation simplifies the momentum dependence:  it is sufficient to consider local self-energies and (spin glass) order parameters so that the replicon operator is easily diagonalized (see Appendix B). The replicon eigenvalue is then related to the so-called spin-glass susceptibility, which is defined in the original disordered model as
\begin{equation}
\chi_{SG}=\frac{1}{V}\int_x\int_y \overline{(\langle \chi(x) \chi(y)\rangle - \langle \chi(x)\rangle \langle\chi(y)\rangle)^2}
\label{SGsusceptibility}
\end{equation}
and is obtained, in principle at least, in measurements on a single macroscopic sample. In this case, the spin-glass susceptibility diverges when the replicon eigenvalue goes to zero (see Appendix B).

The more general setting considered here also encompasses the random field $O(N)$ model as previously discussed   \cite{mezard87,dedominicis95}. The instability associated with the vanishing of the smallest eigenvalue of the replicon operator is harder to interpret than in mean-field-like cases and does not necessarily implies the divergence of the spin-glass susceptibility (see below). It has for instance  been interpreted as signaling the appearance of bound states between pairs of replicas, the replicon operator being then akin to  the corresponding Bethe-Salpeter kernel \cite{brezin01}.

\subsection{Validity of the analytic solution in the explicit RSB formalism}

We now investigate  the stability of the solution of Eq.~(\ref{widetildeG0}) with respect  to the possible appearance of a nonanalytic field dependence when the two field arguments $\phi_{a}$ and $\phi_{b}$ become equal. To this end we apply the operator $\partial_{\phi_{a}} - \partial_{\phi_{b}}$ (with $\partial_{\phi_{a}} \equiv  \partial/\partial \phi_{a}$) to both sides of Eq.~(\ref{widetildeG0}) and use Eqs.~(\ref{widetildeSigma0})  and (\ref{inverse_widetildeG0}). We obtain,    under the natural assumption that  the bare inverse propagator $\widetilde G_0^{[0]-1}$ is regular   when $\phi_{a}\to \phi_{b}$:   
\begin{equation}
\begin{split}
\int_{q'}&\;\; \bigg[\widehat{G}^{[0]}(q;\phi_{a})^{-1} \widehat{G}^{[0]}(q;\phi_{b})^{-1} \delta(q+q') -\\& \frac{\delta \widetilde{\Sigma}^{[0]}(q;\phi_{a},\phi_{b})}{\delta \widetilde{G}^{[0]}(q';\phi_{a},\phi_{b})}\bigg]\left( \partial_{\phi_{a}} - \partial_{\phi_{b}}\right) \widetilde{G}^{[0]}(q';\phi_{a},\phi_{b})\\&= \left( \partial_{\phi_{a}} - \partial_{\phi_{b}}\right) \widetilde{\Sigma}^{[0]}\left(  q;\phi_{a},\phi_{b}\right) \big\vert_{\widetilde{G}^{[0]}(\phi_{a},\phi_{b})} - \\& \widetilde{G}^{[0]}(q;\phi_{a},\phi_{b})\left( \partial_{\phi_{a}} - \partial_{\phi_{b}}\right)\left[\widehat{G}^{[0]}(q;\phi_{a})^{-1} \widehat{G}^{[0]}(q;\phi_{b})^{-1} \right] ,
\end{split}
\label{stability_cusp}
\end{equation}
in which the right-hand side involves derivatives of the self-energy $\widetilde{\Sigma}^{[0]}$ with respect to its explicit dependence on $\phi_{a}$ and $\phi_{b}$ as well as with respect to $\widehat{G}^{[0]}(\phi_{a}),\, \widehat{G}^{[0]}(\phi_{b}),\, \widetilde{G}^{[0]}(\phi_{a},\phi_{a})$ and $\widetilde{G}^{[0]}(\phi_{b},\phi_{b})$, keeping $\widetilde{G}^{[0]}(\phi_{a},\phi_{b})$ {\it  constant}, whereas the derivative of $\widetilde{\Sigma}^{[0]}$ in the left-hand side is taken with $\phi_{a},\phi_{b},\widehat{G}^{[0]}(\phi_{a}),\widehat{G}^{[0]}(\phi_{b}),\widetilde{G}^{[0]}(\phi_{a},\phi_{a}),\widetilde{G}^{[0]}(\phi_{b},\phi_{b})$ constant. We define the operator $\Lambda^{[0]}$ as
\begin{equation}
\begin{split}
&\Lambda^{[0]}(q,q';\phi_{a},\phi_{b}) = \\&\widehat{G}^{[0]}(q;\phi_{a})^{-1} \widehat{G}^{[0]}(q;\phi_{b})^{-1} \delta(q+q')- \frac{\delta \widetilde{\Sigma}^{[0]}(q;\phi_{a},\phi_{b})}{\delta \widetilde{G}^{[0]}(q';\phi_{a},\phi_{b})}.
\end{split}
\label{lambda0}
\end{equation}
Consider now the limit of Eq.~(\ref{stability_cusp}) when $\phi_{a}=\phi_{b}=\phi$. Due to the symmetry properties (recall that the functions are even in $\phi_{a}-\phi_{b}$), the right-hand side of the equation vanishes, as one does not expect in this expression any singular behavior in the field arguments. (In doing so, we are guided by the previous work on nonanalytic field dependence  in disordered systems \cite{fisher86,balents93, chauve98,chauve00,chauve01b, ledoussal02,ledoussal02b,ledoussal03,ledoussal04,ledoussal04b,ledoussal06,ledoussal08b,ledoussal09b,feldman00,feldman01,feldman02,tarjus04,tissier06a,tissier06b,tarjus08,tissier08}  that show that a singularity is indeed only expected in the $\phi_a-\phi_b$ dependence  of  $\widetilde{G}^{[0]}(q';\phi_{a},\phi_{b})$). Eq.~(\ref{stability_cusp}) then becomes
\begin{equation}
\begin{split}
\int_{q'}\Lambda^{[0]}(q,q';\phi,\phi) \left( \partial_{\phi_{a}} - \partial_{\phi_{b}}\right) \widetilde{G}^{[0]}(q';\phi_{a},\phi_{b})\big|_{\phi}=0,
\end{split}
\label{stability_cusp_first}
\end{equation}
where the derivative is evaluated for $\phi_{a}=\phi_{b}=\phi$. If $\Lambda^{[0]}(q,q';\phi,\phi)$ is a strictly positive definite operator, the only solution of the above equation is $(\partial_{\phi_{a}} - \partial_{\phi_{b}}) \widetilde{G}^{[0]}(q';\phi_{a},\phi_{b})\big|_{\phi}=0$. Similarly, all higher-order odd derivatives of $\widetilde{G}^{[0]}$ with respect to $\phi_{a} - \phi_{b}$ (keeping {\it e.g.}  $\phi_{a} + \phi_{b}$ constant) vanish, which characterizes an analytic dependence of the function on $\phi_{a} - \phi_{b}$ near the equality of the 2-replica fields. This is what is expected in the generic paramagnetic phase found at high enough temperature and disorder strength.

On the other hand, if a cusp is present in the field dependence of $\widetilde{G}^{[0]}$, namely, if
\begin{equation}
\begin{array}{ll}
\widetilde{G}^{[0]}(q;\phi_{a},\phi_{b})=\ & \widetilde{G}^{[0]}(q;\phi,\phi)\  +  \\
\\
& \widetilde{G}^{[0]}_{cusp}(q;\phi)\,|\phi_{a} - \phi_{b}|^{\alpha}+\cdots
\end{array}
\label{cusp_G0}
\end{equation}
when $\phi_{a},\phi_{b}\rightarrow \phi$ with $0<\alpha<2$, the operator $\Lambda^{[0]}(q,q';\phi,\phi)$ must have a vanishing eigenvalue so that Eq.~(\ref{stability_cusp_first}) has a solution with $(\partial_{\phi_{a}} - \partial_{\phi_{b}}) \widetilde{G}^{[0]}(q';\phi_{a},\phi_{b})\big|_{\phi}\neq 0$. The same actually applies to any nonanalytic dependence on  $\phi_{a} - \phi_{b}$ since it implies that a given odd derivative of the function (under the action of  $\partial_{\phi_{a}} - \partial_{\phi_{b}}$) is nonzero despite the fact that the function is even in $\phi_{a} - \phi_{b}$.  Conversely,   if  the operator  $\Lambda^{[0]}(q,q';\phi,\phi)$  becomes marginal with a vanishing  eigenvalue, the function $\widetilde{G}^{[0]}$ may be written in the vicinity of $\phi_{a}=\phi_{b}=\phi$ as
\begin{equation}
\begin{array}{ll}
\widetilde{G}^{[0]}(q;\phi_{a},\phi_{b})=\ &\widetilde{G}^{[0]}(q;\phi,\phi) + F(\phi_{a} - \phi_{b})\, \Psi_0(q;\phi)\  +  \\ 
\\
& \Psi_{\bot}(q;\phi_{a}, \phi_{b}),
\label{nonanalytic_G0}
\end{array}
\end{equation}
where $\Psi_0(q;\phi)$ is the eigenvector associated with the vanishing eigenvalue of $\Lambda^{[0]}(q,q';\phi,\phi)$, $F(\phi_{a} - \phi_{b})$ is an even and possibly nonanalytic function of its argument, and $\Psi_{\bot}$ is orthogonal to $\Psi_0(q;\phi)$, regular in its field arguments and even in $\phi_{a} - \phi_{b}$ . It is easy to verify that the above expression Eq.~(\ref{nonanalytic_G0})  satisfies Eq.~(\ref{stability_cusp_first}) and all higher-order equations involving odd derivatives. The solution must of course satisfy Eq.~(\ref{stability_cusp}), which puts constraints on the functional form of $F(\phi_{a} - \phi_{b})$. One can however wonder if the Schwinger-Dyson equation Eq.~(\ref{widetildeG0}) combined with Eq.~(\ref{widetildeSigma0})  is still valid when a cusp in $|\phi_{a} - \phi_{b}|$ has appeared in $\widetilde{G}^{[0]}$.   This question has been examined in detail in the large $N$ version of the random manifold model at leading order \cite{ledoussal02,ledoussal03,ledoussal08b}    where it has been shown that a cusp indeed occurs.    In this case,  the analog of Eqs.~(\ref{widetildeG0},\ref{widetildeSigma0}) have no solution in the ``cuspy'' region.  However,  the main point of the analysis of [\onlinecite{ledoussal02,ledoussal03,ledoussal08b}]  is that when put in a \textit{differential} form, via a FRG approach,  the equations can then be continued and solved.  

\subsection{Proof of equivalence}
To prove the identity of the operator $\Lambda^{[0]}(q,q';\phi,\phi)$ signaling the limit of validity of the analytic solution in an explicit RSB scheme and the replicon operator $\Lambda_R(q,q';\phi)$   of the RS scheme, it is convenient to rederive $\Lambda^{[0]}$ by applying the operator $\partial_{\phi_{a}} - \partial_{\phi_{b}}$ to the general Schwinger-Dyson equation, Eqs. (\ref{saddlepoint},\ref{GG}), and by using an expansion in free replica sums only at a later stage of the process. 

Applying the operator ${\partial_{\phi_{a}}}-{\partial_{\phi_{b}}}$ to Eq.~(\ref{saddlepoint}) with $a\neq b$ and with uniform sources leads to 
\begin{equation}
\begin{array}{ll}
\displaystyle 0 = \left({\partial_{\phi_{a}}}-{\partial_{\phi_{b}}}\right){\delta \Gamma_{2PI} [ \{\phi_{e}\}, \{G_{ef}\}]\over \delta G_{ab}(q)}\bigg|_{*}\;+\\
\\
\displaystyle \sum_{c,d}\int_{q'} \ {\delta^2 \Gamma_{2PI}[ \{\phi_{e}\}, \{G_{ef}\}]\over \delta G_{ab}(q)\delta G_{cd}(q')}\bigg\vert_{*}\ \left({\partial_{\phi_{a}}}-{\partial_{\phi_{b}}}\right) G^*_{cd}(q';\{\phi_{e}\})
\label{derivGamma2}
\end{array}
\end{equation}
where we have momentarily reintroduced a star to indicate that the quantities are evaluated with the solution of the Schwinger-Dyson equations (symbolically, $G=G^*$). 

We first study the first term of the right-hand side of Eq.~(\ref{derivGamma2}). From Eqs.~(\ref{Gamma2PI}) and  (\ref{selfsigma}) one has
\begin{equation}
\begin{array}{ll}
\displaystyle  2 {\delta \Gamma_{2PI} [\{\phi_{e}\}, \{G_{ef}\}]\over \delta G_{ab}(q)}= \\
\\
\hspace{1cm} - G^{-1}_{ab}(q)+G^{-1}_{0 ab}(q;\{\phi_{e}\})+\Sigma_{ab}(q;\{\phi_{e}\},\{G_{ef}\})\  .
\label{dergam}
\end{array}
\end{equation}
We apply ${\partial_{\phi_{a}}}-{\partial_{\phi_{b}}}$ to this expression and evaluate the result both at the solution $G=G^*$ and in a configuration where all the replica fields $\phi_{e}$  are equal to $\phi$.
Under this operation the first term in the right-hand member of  Eq.~(\ref{dergam})  trivially vanishes  since the $G_{ef}$'s and the $\phi_{e}$'s are considered here as independent variables.  The second and third terms also vanish under the assumption that the dependence of  $G^{-1}_{0 ab}(q;\{\phi_{e}\})$ and $\Sigma_{ab}(q;\{\phi_{e}\}, \{G_{ef}\})$  on the $\phi_{e}$'s (at \textit{constant} $G_{ef}$'s) is regular: the two functions being symmetric in the exchange of the two fields  $\phi_{a}$ and $\phi_{b}$, the difference of derivatives vanish in the limit where all replica fields become equal.

We now  consider the second term of the right-hand side of Eq.~(\ref{derivGamma2}), which we also evaluate in a configuration with all replica fields equal to $\phi$. Introducing 
\begin{equation}
\begin{split}
M_1(q,q')&=2 \, \frac{\delta^2 \Gamma_{2PI}[ \{\phi_{e}\}, \{G_{ef}\}]}{\delta G_{ab}(q)\delta G_{aa}(q')}\bigg\vert_{*,\phi}\\&= 2 \, \frac{\delta^2 \Gamma_{2PI}[ \{\phi_{e}\}, \{G_{ef}\}]}{\delta G_{ab}(q)\delta G_{bb}(q')}\bigg\vert_{*,\phi}
\end{split}
\end{equation}
and, when $c\neq a,b$,
\begin{equation}
M_2(q,q')=2 \, \frac{\delta^2 \Gamma_{2PI}[ \{\phi_{e}\}, \{G_{ef}\}]}{\delta G_{ab}(q)\delta G_{cc}(q')}\bigg\vert_{*,\phi},
\end{equation}
we can thus rewrite Eq.~(\ref{derivGamma2}) as 
\begin{equation}
\begin{array}{ll}
\displaystyle \int_{q'} \ \bigg\{M_1(q,q')  \left({\partial_{\phi_{a}}}-{\partial_{\phi_{b}}}\right) \left[ G^*_{aa}(q';\{\phi_{e}\})+G^*_{bb}(q';\{\phi_{e}\})\right]\big\vert_{\phi}\\
\\
\displaystyle + M_2(q,q') \sum_{c\ne a,b} \left({\partial_{\phi_{a}}}-{\partial_{\phi_{b}}}\right) G^*_{cc}(q';\{\phi_{e}\})\big\vert_{\phi}\bigg\} \\
\\
\displaystyle + \int_{q'}    \big({\partial_{\phi_{a}}}-{\partial_{\phi_{b}}}\big) \bigg\{2 \Lambda_1(q,q')\, G^*_{ab}(q';\{\phi_{e}\})\\
\\
\displaystyle  \  + 2 \Lambda_2(q,q')\sum_{c\ne a,b}\bigg[G^*_{ac}(q';\{\phi_{e}\}) + G^*_{bc}(q';\{\phi_{e}\})\bigg]\\
\\
\displaystyle  +\Lambda_3(q,q') \sum_{\substack{c,d\ne a,b\\ c\ne d}} G^*_{cd}(q';\{\phi_{e}\}) \bigg\}\bigg\vert_{\phi}=0
\label{derivGamma2plus}
\end{array}
\end{equation}
where the $\Lambda_i$'s, $i=1,2,3$, are evaluated in a configuration with all replica fields equal (\textit{i.e.}, replica symmetric) and are given by Eqs.~(\ref{matrixM},\ref{Lamdda_i's}).  They can then be moved under the action of the operator ${\partial_{\phi_{a}}}-{\partial_{\phi_{b}}}$. The second term of the left-hand side of  Eq.~(\ref{derivGamma2plus}) can be further reexpressed as
\begin{equation}
\begin{array}{ll}
\displaystyle \int_{q'}   \bigg\{2\left(\Lambda_1-2 \Lambda_2+\Lambda_3\right) \, \left({\partial_{\phi_{a}}}-{\partial_{\phi_{b}}}\right)G^*_{ab}(q';\{\phi_{e}\}) \big\vert_{\phi}\,+ \\
\\
\displaystyle 2(\Lambda_2-\Lambda_3) \, \left({\partial_{\phi_{a}}}-{\partial_{\phi_{b}}}\right)\bigg[\sum_{c \ne a}G^*_{ac}(q';\{\phi_{e}\}) \,+ \\
\\
\displaystyle \sum_{c \ne b}G^*_{bc}(q';\{\phi_{e}\})\bigg]\bigg\vert_{\phi}+\Lambda_3  \left({\partial_{\phi_{a}}}-{\partial_{\phi_{b}}}\right) \sum_{c\ne d}G^*_{cd}(q';\{\phi_{e}\})\bigg\vert_{\phi}\bigg\}. 
\label{derivgamma2organ}
\end{array}
\end{equation}
We finally expand the different correlation functions $G^*_{ab}(q;\{\phi_{e}\})$ in number of free replica sums. Under weak assumptions, namely that all first derivatives are bounded and that no singularities are encountered in the free sums over replicas, \textit{e.g.} that
\begin{equation}
\partial_{\phi_{a}}\sum_{c}\widetilde{G}^{[0]}(q;\phi_{a},\phi_{c})]\big\vert_{\phi}=\partial_{\phi_{b}}\sum_{c}\widetilde{G}^{[0]}(q;\phi_{b},\phi_{c})]\big\vert_{\phi},
\end{equation}
one readily finds that
\begin{equation}
\begin{split}
\left({\partial_{\phi_{a}}}-{\partial_{\phi_{b}}}\right) \bigg[ G^*_{aa}(q;\{\phi_{e}\})+G^*_{bb}(q;\{\phi_{e}\})\bigg]\bigg\vert_{\phi}=0,\\&
\\
\left({\partial_{\phi_{a}}}-{\partial_{\phi_{b}}}\right)  G^*_{cc}(q;\{\phi_{e}\})\big\vert_{\phi}=0, \, \mathrm{when}\, c\ne a,b,
\end{split}
\end{equation}
so that  the first term in the left-hand side of    Eq.~(\ref{derivGamma2plus}) vanishes
and  the only surviving term in Eq.~(\ref{derivgamma2organ}) is the one involving the combination $\Lambda_1-2 \Lambda_2+\Lambda_3$. As a final result, Eq.~(\ref{derivGamma2plus}) then reduces to Eq.~(\ref{stability_cusp_first}) with
\begin{equation}
\begin{split}
\Lambda^{[0]}(q,q';\phi, \phi)&=\Lambda_1(q,q';\phi)-2\Lambda_2(q,q';\phi)+\Lambda_3(q,q';\phi)\\
\\
&=\Lambda_R(q,q';\phi).
\end{split}
\label{result_replicon}
\end{equation}
The $\Lambda_i$'s are equivalently obtained as zeroth-order contributions in the free replica sum expansions for a configuration of equal replica fields or from the RS solution when $n\rightarrow 0$ (the two approaches being equivalent as discussed in section III). Eq.~(\ref{result_replicon}) is our main result: it proves the equivalence of the operator signaling a breakdown of the analytic solution in the explicit RSB scheme and the replicon operator associated with the stability of the RS solution in the standard replica scheme.  The demonstration can be easily extended to systems described by an $N$-component field.

In Appendix A, we consider the random field $O(N)$ model  in the 2PI formalism at leading and next-to-leading orders of the $1/N$ expansion. We explicitly show that the operator associated with the instability (or breakdown) of the analytic solution in the explicit RSB scheme reduces to the replicon operator derived in Refs.~[\onlinecite{mezard94, dedominicis95}] in the appropriate limit.

\section{Discussion}

\subsection{Susceptibilities and spin-glass behavior}
As pointed out in Ref.~[\onlinecite{dedominicis95}], the inverse of the replicon operator, $\chi_R(q,q')=\Lambda_R^{-1}(q,q')$, plays the role of a susceptibility (operator), and a continuous ``transition'' associated with spontaneous RSB is characterized by the divergence of the largest eigenvalue of the susceptibility operator. Similarly, according to the arguments of section IV, such a divergence also signals the appearance of a nonanalytic field dependence of  2-replica correlation functions. It is instructive to relate this susceptibility operator to other quantities. On the one hand, $\chi_R(q,q')$ is connected to the standard spin-glass susceptibility, which is defined in Eq.~(\ref{SGsusceptibility}), as \cite{dedominicis95}
\begin{equation}
\chi_{SG}= \int_q \int_{q'}\chi_R(q,q').
\label{SG_chiR}
\end{equation}
\\

On the other hand, $\chi_R(q,q')$ can be expressed in terms of usual many-point correlation (Green) functions that are generated by  derivatives of the replicated free-energy functional $W_{rep}$ with respect to the sources $J_a(x)$, in the absence of any source $K_{ab}(x,y)$ coupled to composite replica fields. After introducing
\begin{equation}
W_{abcd}^{(4)}(q,-q,q',q'';\phi) = \delta(q'+q'')\ G_{abcd}^{(4)}(q,q';\phi),
\end{equation}
where $W_{abcd}^{(4)}$ is the 4th derivative of the generating functional $W$ with respect to $J_a,J_b,J_c,J_d$ and is evaluated in a RS configuration with all replica fields equal to $\phi$, and \cite{tissier08}
\begin{equation}
\begin{split}
&G_R^{(4)}(q,q';\phi) = \\ &
\\
&G_{abab}^{(4)}(q,q';\phi) -2 G_{abac}^{(4)}(q,q';\phi) + G_{abcd}^{(4)}(q,q';\phi),
\end{split}
\end{equation}
where distinct Latin indices denote here distinct replicas, one finds that
\begin{equation}
\chi_R(q,q') = G_R^{(4)}(q,q';\phi) + \delta(q+q')\, \widehat{G}(q;\phi)^2,
\end{equation}
with, we recall, $\widehat{G}(q;\phi)=G_C(q;\phi)$ (see section III). If one expresses the 2-replica propagator in the explicit RSB scheme at zeroth order, $\widetilde{G}^{[0]}(q)$, not as a function of the replica fields $\phi_a$, $\phi_b$ but as a function of the (uniform) sources $J_a$, $J_b$, one also derives that
\begin{equation}
\begin{split}
\partial_{J_a}\partial_{J_b}\widetilde{G}^{[0]}(q; J_a, J_b)\big |_{J} &= G_R^{(4)}(q,0;\phi) =G_R^{(4)}(0,q;\phi) \\ &
\\
&= \chi_R(q,0) - \delta(q)\,  \widehat{G}(q;\phi)^2,
\end{split}
\end{equation}
where the partial derivative is evaluated for $J_a = J_b =J$. From this, it can be seen that if the passage from $\phi$ to $J$ is nonsingular, \textit{i.e.} if the ferromagnetic susceptibility $\chi_{FM}=\widehat{G}(q=0)$ stays finite, the existence of a cusp in  $\phi_a - \phi_b$, or alternatively in $J_a - J_b$, implies a divergence of $\chi_R(q,0)=\chi_R(0,q)$.

This exercise may prove interesting in the case of disordered systems for which it can be shown that the spin-glass susceptibility is bound from above by the ferromagnetic susceptibility, $\chi_{SG} \leq \chi_{FM}$, and therefore cannot diverge if the latter stays finite. This property has recently been shown for the random field Ising model \cite{krzakala09}. In such cases, no spin-glass phase, defined as a phase with a diverging $\chi_{SG}$, can be found outside a paramagnetic-ferromagnetic critical point. However, this argument does not preclude the existence of a phase with spontaneous RSB and/or a phase with cuspy 2-replica correlation functions. What is required for such a phase is the divergence of the largest value of the replicon susceptibility \textit{operator}. If this eigenvalue is of measure zero in the spectrum, its divergence may not necessarily trigger that of $\chi_{SG}$ (see Eq.~(\ref{SG_chiR})). On the other hand, it could still imply the divergence of $\chi_R(q,0)$. This is admittedly a quite intricate scenario. Calling ``glassy'' such a phase with anomalous behavior but no diverging $\chi_{SG}$ has anyway little physical motivation unless the glassy characteristics are supported by other phenomena such as slow dynamics.

\subsection{Renormalization group}
Formally relating the stability conditions of the RS solution in the standard replica method and of the analytic solution in the explicit RSB scheme is of course only one piece of the puzzle. It remains to be seen for specific disordered systems if RSB is actually spontaneously broken and if a cusp actually appears in the field dependence of the 2-replica pair correlation functions. One potential advantage of the explicit RSB scheme is that it can be combined with renormalization group ideas, more specifically with the FRG. In disordered systems such as random manifold  \cite{fisher85,fisher86,balents93, chauve98,chauve00,chauve01b, ledoussal02b,ledoussal04,ledoussal02,ledoussal03,ledoussal08b,balents05,ledoussal09b}, random field and random anisotropy \cite{fisher85,feldman00,feldman01,feldman02, tarjus04,tissier06a,tissier06b,tarjus08,tissier08,ledoussal06} models, it has indeed been possible to generate an infinite hierarchy of flow equations for the renormalized cumulants of the disorder (which are related to the many-replica correlation functions for \textit{nonequal} sources \cite{tarjus08}). On the other hand, implementing RG schemes to take into account fluctuations and go beyond mean-field treatment has proven difficult in the presence of \textit{spontaneous} RSB (see \textit{e.g.} Ref.[\onlinecite{dedominicis98,dedominicis06}]), which, for instance, still prevents one from reaching definite conclusions concerning finite-dimensional spin glasses. Looking for cuspy behavior in an FRG formalism with explicit RSB therefore appears as a promising alternative.

The presentation given in the previous sections of the explicit RSB method within the 2PI formalism has however not incorporated any RG scheme. RG can be implemented through the introduction of appropriate cut-off functions that suppress the contribution of the low-momentum modes below some running scale $k$ \cite{wilson74, polchinski84, wetterich93c, berges02,pawlowski07,delamotte03}. This, as well as the connection with the 1PI FRG formalism developed in Refs. [\onlinecite{tarjus04,tissier06a,tissier06b,tarjus08,tissier08}], will be discussed in a forthcoming publication \cite{mouhanna10}. For now, we note that quite generally a cusp is associated with ``shocks'' or ``avalanches'' \cite{balents96,balents05,ledoussal09b,tissier08} which correspond to abrupt changes in the ground state of the system. (We focus here on equilibrium conditions, hence on the ground state, not on metastable states and hysteresis phenomena \cite{sethna07}). At  zero temperature, for many disordered systems (random field and anisotropy, spin glass, etc...), such abrupt changes are generic (see \textit{e.g.} the simulation results in Refs.~[\onlinecite{frontera00,wu05, liu07}]) so that one indeed expects to observe cusp behavior in the field dependence of the $2$-replica correlation functions. However, important questions are then whether such a phenomenon survives at long distances, \textit{e.g.} at the critical point in random-field systems, and if it persists at nonzero temperature where finite-scale discontinuities are usually believed to be rounded. FRG is a tool to answer such questions.\footnote{For instance, it would be interesting to reconcile the claim made by Mezard et al. \cite{mezard92, mezard94}  and De  Dominicis et al.  \cite{dedominicis95} that the replicon operator of the RF$O(N)$ model in the $1/N$ expansion at next-to-leading order becomes negative even outside the critical point whereas both perturbative (in $d=4+\epsilon$) and nonperturbative FRG results \cite{tarjus04,tissier06a,tissier06b,tarjus08,tissier08}  show that no linear cusp appears in the large $N$ regime of the model.}

\appendix
\section{1/N expansion for the random field O(N) model}

The $1/N$ expansion of the 2PI effective action has been  developed mainly in the context of ouf-of-equilibrium dynamics  to derive  nonperturbative approximations.  The rules defining this  expansion have been dicussed at length in Ref.~[\onlinecite{cornwall73,aarts02,alford04}]. We only recall here the main features of this expansion and we closely follows the notations employed in  Ref.~[\onlinecite{aarts02,alford04}].

The action is  assumed to be of order $N$ and the  fields of order $\sqrt N$.  The 2PI effective action is a singlet under the rotation group $O(N)$.  It must be constructed from the $O(N)$ invariants: $\pmb \phi^2$,  $\hbox{Tr}\,\pmb{G}^p$  and $\hbox{Tr}\, \pmb\phi^2 \pmb{G}^p$. We write the  $1/N$ expansion of $\Gamma_2[ \pmb{\phi}, \pmb{G}]$, which is the sum of 2PI diagrams in the expression of $\Gamma_{2PI}$ (see Eq.~(\ref{Gamma2PI})),  as
\begin{equation}
\Gamma_2[ \pmb{\phi}, \pmb{G}]=\Gamma_2^{LO}[ \pmb{\phi}, \pmb{G}]+ \Gamma_2^{NLO}[ \pmb{\phi}, \pmb{G}]\ +\dots \ . 
\end{equation}
where $\Gamma_2^{LO}[ \pmb{\phi}, \pmb{G}]$ and   $\Gamma_2^{NLO}[ \pmb{\phi}, \pmb{G}]$ correspond to the leading order and next to leading contributions that gather terms of order $N$ and $1$, respectively. 

We start  with the replicated action for the random field O(N) model \cite{mezard92}:
\begin{equation}
\begin{array}{ll}
S[\{\pmb \chi_{a}\}]\displaystyle &=\displaystyle \int_x \bigg\{{1\over 2} \sum_{a=1}^n \left( (\partial \pmb\chi_{a})^2 + m^2 \vert \pmb \chi_{a} \vert ^2 + {\lambda\over 12N}  \vert \pmb \chi_{a} \vert ^4\right)
\\
\\
&\displaystyle  -\sum_{a,b =1}^n  {\Delta\over 2}\   \pmb \chi_{a} .\pmb \chi_{b} \bigg\},
\end{array}
\end{equation}
where $\pmb\chi_{a}$ is an $N$-component replica  field  with  $a=1...n$ and $\Delta$ is the bare second cumulant of the (Gaussian distributed) random field.  Latin indices are used for replicas and Greek indices for vector components.

For uniform replica field configurations $\pmb\phi_{a}$, the free inverse propagator is then given in momentum space by
\begin{equation}
\begin{array}{ll}
[G_0^{-1}]^{\mu \nu}_{ab}(q, q')&=\displaystyle \bigg\{\bigg({q}^2+m^2+{\lambda\over 6N}   \vert \pmb \phi_{a} \vert ^2\bigg)\,  \delta^{\mu \nu} \delta_{ab}\\
\\
&\displaystyle +{\lambda\over 3N} \phi_{a}^{\mu} \phi_{a}^{\nu} \delta_{ab}-\Delta \delta^{\mu \nu} \bigg\}  \delta(q+q')\\
\\
&\displaystyle \equiv  [G_0^{-1}]^{\mu \nu}_{ab}(q)\ \delta(q+q'),
\label{propagfourier}
\end{array}
\end{equation}
where  $[G_0]^{\mu \nu}_{a b}$  are the components of $\pmb{G}_{0,ab}$.

The leading-order (LO) contribution to $\Gamma_2[ \pmb{\phi}, \pmb{G}]$  is simply obtained from its counterpart in the  $O(N)$ case, up to the introduction of replica indices:
\begin{equation}
\Gamma_2^{\scriptsize\hbox{LO}}[ \pmb{\phi}, \pmb{G}]=\displaystyle{\lambda\over  4! N} \sum_{a} \int_x \left(  \sum_{\mu} G^{\mu \mu}_{a a}(x,x)\right) ^2. 
\end{equation} 
The above expression has a graphical interpretation which is a direct extension of what is used in the standard $O(N)$ model. Although graphs provide a compact representation, we shall not give them in this appendix. 
 
The next-to-leading  order (NLO) is also given by a generalization of the $O(N)$ case. To derive it,  it is convenient to consider separately  the $\pmb{\phi}$-independent part  $\Gamma_{2,ind}^{\scriptsize\hbox{NLO}}$  and the  $\pmb{\phi}$-dependent one $\Gamma_{2,dep}^{\scriptsize\hbox{NLO}}$\,. The former is generated by the   vertex associated with   $\chi_a^{\mu} \chi_a^{\mu}  \chi_a^{\nu}\chi_a^{\nu} $. This contribution can be formally written as
\begin{equation}
\Gamma_{2,ind}^{\scriptsize\hbox{NLO}}[\pmb{G}]=\displaystyle{1\over 2}  \hbox{Tr}\ln (B)
\label{NLOchampnul}
\end{equation}
with
\begin{equation}
\hbox{B}_{ab}(x,y;\pmb{G})=\displaystyle \delta(x-y) \delta_{ab}+{\lambda\over  6 N} \sum_{\mu,\nu}G^{\mu \nu}_{a b}(x,y)\, G^{\mu \nu}_{a b}(x,y)
\label{definitionB}
\end{equation}
and where the trace acts both on replica indices and  spatial  coordinates.

In turn, the   $\pmb{\phi}$-dependent part of the NLO contribution is generated by the vertex associated with $\phi_a^{\mu} \chi_a^{\mu}  \chi_b^{\nu}\chi_b^{\nu} $ with $\phi_a^{\mu}$  being considered as an inserted operator. It is given by 
\begin{equation}
\begin{array}{ll}
\Gamma_{2,dep}^{\scriptsize\hbox{NLO}}[\pmb\phi,\pmb{G}]=\\
\\\displaystyle  -{\lambda\over  6 N} \int_x \int_y  \sum_{a,b} \hbox{I}_{ab}(x,y;\pmb{G}) \sum_{\mu,\nu}{\phi}_a^{\mu}(x)\, G^{\mu \nu}_{a b}(x,y)\,\phi_b^{\nu} (y)
\label{NLOphinonul}
\end{array}
\end{equation}\\
with $\hbox{I}_{ab}(x,y;\pmb{G})$ iteratively defined as
\begin{equation}
\begin{array}{ll}
&\hbox{I}_{ab}(x,y;\pmb{G})=\displaystyle{\lambda\over  6 N} \sum_{\mu,\nu}G^{\mu \nu}_{a b}(x,y)\, G^{\mu \nu}_{a b}(x,y)\\
\\
&\displaystyle -{\lambda\over  6 N}\int_z \sum_{c} \  \hbox{I}_{ac}(x,z;\pmb{G})\sum_{\mu,\nu}G^{\mu \nu}_{c b}(z,y)\, G^{\mu \nu}_{c b}(z,y)\ . 
\label{definitionI}
\end{array}
\end{equation}
Note that $\hbox{I}_{ab}(x,y;\pmb{G})$ and $\hbox{B}_{ab}(x,y;\pmb{G})$ are related  by
\begin{equation}
 \hbox{B}^{-1} _{ab}(x,y;\pmb{G})=\delta(x-y) \delta_{ab}-\hbox{I}_{ab}(x,y;\pmb{G}).
\label{relationIB}
\end{equation}

Finally, gathering all LO and NLO  contributions and considering uniform replica field configurations leads to
\begin{equation}
\begin{array}{ll}
\displaystyle{\Gamma_2[ \pmb{\phi}, \pmb{G}]\over V}&= \displaystyle{\lambda\over  4! N}    \sum_{a}  \left(\int_q \sum_{\mu} G^{\mu\mu}_{a a}(q)\right)^2 \\
\\
&\displaystyle +{1\over 2} \sum_{a}\int_q \ \left[ \ln(B)\right] _{aa}(q)\\
\\
&\displaystyle-{\lambda\over  6 N}\   \sum_{a,b} \sum_{\mu,\nu} \phi^{\mu}_{a} \phi^{\nu}_{b} \int_q\  \hbox{I}_{ab}(q;\pmb{G})\ G^{\mu \nu}_{a b}(q).
\label{defGamma2p}
\end{array}
\end{equation}

It is convenient to introduce a ``polarization''
\begin{equation}
\Pi_{ab}(q;\pmb{G})=\displaystyle{\lambda\over  6 N}  \int_{q'}\   \sum_{\mu,\nu} G^{\mu \nu}_{a b}(q')\, G^{\mu \nu}_{a b}(q+q')
\label{definitionPi}
\end{equation}
and a quantity \cite{mezard92}
\begin{equation}
V_{ab}(q;\pmb{G})=\delta_{ab}-\hbox{I}_{ab}(q;\pmb{G})= \left[ \nbOne+\Pi(q;\pmb{G})\right] ^{-1}_{ab},
\end{equation}
from which $B$ and $I$ can be simply expressed as
\begin{equation}
\hbox{B}_{ab}(q,q';\pmb{G})]=\delta(q+q')\left[ \delta_{ab}+ \Pi_{ab}(q;\pmb{G})\right] ,
\end{equation}
\begin{equation}
\hbox{I}_{ab}(q;\pmb{G})=\sum_{c} \Pi_{ac}(q;\pmb{G})\ V_{cb}(q;\pmb{G}). 
\end{equation}

\subsection{Schwinger-Dyson equation}
We now derive the Schwinger-Dyson equation for the propagator. To this end we have to compute the  functional derivative 
$\delta  \Gamma_2[\pmb{\phi},\pmb{G}]/ \delta G^{\mu \nu}_{ab}(x,y)$. Since we only consider  uniform field configurations, it is simpler to work in momentum space, which provides
\begin{equation}
\begin{array}{ll}
\displaystyle{\delta  \Gamma_2[\pmb{\phi},\pmb{G}]\over \delta  G^{\mu \nu}_{ab}(q)}=\displaystyle{\lambda\over  12 N}\, \delta^{\mu \nu}\,    \delta_{ab} \int_{q'}  \sum_{\mu}  G^{\mu \mu}_{aa}(q') \\
\\
\displaystyle +{\lambda \over 6N}\int_{q'}\ V_{ba}(q';\pmb{G})\, G^{\mu \nu}_{ab}(q+q')-{\lambda\over  6 N}\ \displaystyle \hbox{I}_{ab}(q;\pmb{G})\,\phi_a^{\mu}\phi_b^{\nu}\\
\\
\displaystyle-{1\over 2}{\left(\lambda \over  3 N\right)^2}\int_{q'} \sum_{c,d;\rho,\sigma}  \phi_c^{\rho}\, \phi_d^{\sigma}\, V_{ca}(q';\pmb{G})\, G^{\mu \nu}_{ab}(q+q') \times \\
\\
\displaystyle \times V_{bd}(q';\pmb{G})\,  G^{\sigma \rho}_{dc}(q').
\label{derivegamma2}
\end{array}
\end{equation}
Thus, the Schwinger-Dyson equation for the propagator,  Eqs.~(\ref{GG},\ref{selfsigma}), can be written in terms of  vectors and tensors of $O(N)$ as
\begin{equation}
\begin{array}{ll}
&\displaystyle [\pmb G^{-1}(q)]_{ab}-[\pmb G^{-1}_0(q)]_{ab}=\displaystyle{\lambda\over  6 N}\, \pmb\nbOne_N \,    \delta_{ab}   \int_{q'}\  \hbox{Tr}_{O(N)}\  \pmb G_{aa}(q') \\
\\
&\displaystyle +{\lambda\over 3N}\int_{q'}  V_{ba}(q';\pmb{G})\, \pmb G_{ab}(q+q')-{\lambda\over  3 N}\ \displaystyle \hbox{I}_{ab}(q;\pmb{G})\,\pmb\phi_{a}\,  \pmb \phi_b^T \\
\\
&\displaystyle-{\left(\lambda\over  3N\right)^2}\int _{q'} \sum_{c,d} \, V_{ca}(q';\pmb{G})V_{bd}(q';\pmb{G})\, \pmb G_{ab}(q+q') \times \\
\\
 &\displaystyle \times \, \pmb \phi_d \, [\pmb G_{dc}(q')\, \pmb\phi_{c}]^T.
\label{derivegamma2}
\end{array}
\end{equation}
where $\pmb\phi_{a}$ is considered as a column vector and the superscript $T$ denotes a transposed quantity. For further convenience, we introduce the function $\pmb C_{ab}(q)=[\pmb G^{-1}(q)]_{ab}$ which is nothing but the 2-point proper vertex. The above equations are generalizations to nonzero sources of those derived by Mezard and Young \cite{mezard92} within Bray's self-consistent screening approximation \cite{bray74}, which is a zero-source version of the $1/N$ expansion of the 2PI formalism.

\subsection{Expansion in free replica sums}
We now expand the different functions entering in (\ref{derivegamma2})  in increasing number of free replica sums.  The general principles of the method are presented in section III. The functions $\pmb G_{ab}(q), \pmb C_{ab}(q)$,  $\Pi_{ab}(q), V_{ab}(q), I_{ab}(q)$ are decomposed as in Eq.~(22) and their ``hat'' and ``tilde'' components are expanded as in Eq.~(23) and (24).

For the zeroth-order terms, one easily derives the following relations:
\begin{equation}
\begin{array}{ll}
\widehat{\pmb G}^{[0]}(q;\pmb\phi_a)& =\  {\widehat{\pmb C}^{[0]-1}}(q;\pmb\phi_a) ,\\
\\
\widetilde{\pmb G}^{[0]}(q;\pmb\phi_a,\pmb\phi_b)& =- \ {\widehat{\pmb G}^{[0]}}(q;\pmb\phi_a)\   \widetilde{\pmb C}^{[0]}(q;\pmb\phi_a,\pmb\phi_b)\ {\widehat{\pmb G}^{[0]}}(q;\pmb\phi_b),
\end{array}
\label{decompoG}\end{equation}
\begin{equation}
\begin{array}{ll}
\widehat{V}^{[0]}(q;\pmb\phi_a)& =\  [1+{\widehat{\Pi}^{[0]}(q;\pmb\phi_a)]^{-1}}, \\
\\
\widetilde{V}^{[0]}(q;\pmb\phi_a,\pmb\phi_b)& =- \widehat{V}^{[0]}(q; \pmb\phi_a)\   \widetilde{\Pi}^{[0]}(q;\pmb\phi_a,\pmb\phi_b)\   \widehat{V}^{[0]}(q;\pmb\phi_b), 
\end{array}
\label{decompoV}
\end{equation}
\begin{equation}
\begin{array}{ll}
\widehat{\hbox{I}}^{[0]}(q;\pmb\phi_a) &=\  \widehat{\Pi}^{[0]}(q;\pmb\phi_a)\ \widehat{V}^{[0]}(q;\pmb\phi_a), \\
\\
\widetilde{\hbox{I}}^{[0]}(q;\pmb\phi_a,\pmb\phi_b) &=\ -\widetilde{V}^{[0]}(q;\pmb\phi_a,\pmb\phi_b),
\end{array}
\label{decompoI}
\end{equation}
while  from the definition in Eq.~(\ref{definitionPi}), the components of the polarization can be expressed as
\begin{equation}
\begin{array}{ll}
&\displaystyle \widehat{\Pi}^{[0]}(q;\pmb\phi_a) = {\lambda\over 6N} \int_{q'}\  \hbox{Tr}_{O(N)} \bigg\{\widehat{\pmb G}^{[0]}(q';\pmb\phi_a) \times \\
\\
&\displaystyle   \left( \widehat{\pmb G}^{[0]}(q+q';\pmb\phi_a)  + 2\ \widetilde{\pmb G}^{[0]}(q+q';\pmb\phi_a,\pmb\phi_a)\right) \bigg\}
\end{array}
\label{hatPi}
\end{equation}
and
\begin{equation}
\begin{array}{ll}
&\widetilde{\Pi}^{[0]}(q;\pmb\phi_a,\pmb\phi_b)=\\
\\
&\displaystyle \ {\lambda\over 6N} \int_{q'}\  \hbox{Tr}_{O(N)} \left\{\widetilde{\pmb G}^{[0]}(q';\pmb\phi_a,\pmb\phi_b) \  \widetilde{\pmb G}^{[0]}(q+q';\pmb\phi_a,\pmb\phi_b) \right\rbrace.
\end{array}
\label{tildePi}
\end{equation}
\\

Next, Eqs.~(\ref{decompoG})-(\ref{tildePi}) are inserted in Eq.~(\ref{derivegamma2}) with a separate treatment of the diagonal and nondiagonal components of the replica matrices. One thus obtains two coupled self-consistent equations for  $\widehat{\pmb C}^{[0]}(q;\pmb \phi_a)=\widehat{\pmb G}^{[0]-1}(q;\pmb\phi_a)$ and $\widetilde{\pmb C}^{[0]}(q;\pmb\phi_a,\pmb\phi_b)$.

To proceed further, one must derive the explicit expressions for the ``bare'' or free quantities  $\widehat{\pmb C}_0^{[0]}(q;\pmb\phi_a)=\widehat{\pmb G}_0^{[0]-1}(q;\pmb\phi_a)$ and $\widetilde{\pmb C_0}^{[0]}(q;\pmb\phi_a,\pmb\phi_b)$.  From Eq.~(\ref{propagfourier}), one finds
\begin{equation}
\begin{array}{ll}
\displaystyle\widehat{C}_{0\, ab}^{[0] \mu \nu}(q;\pmb\phi_a)=\bigg[{q}^2+m^2+{\lambda\over 6N}  \vert \pmb \phi_{a}  \vert ^2\bigg]\,  \delta^{\mu \nu} +{\lambda\over 3N} \phi_a^{\mu} \phi_b^{\nu},\\
\\
\widetilde{C}_{0\, ab}^{[0] \mu \nu}(q;\pmb\phi_a,\pmb\phi_b)=- \Delta\ \delta^{\mu \nu}\  .
\end{array}
\end{equation}

We first consider the $1$-replica quantities (\textit{i.e.} quantities depending on a single replica field). For explicit calculations, one   specifies a particular configuration for  $\pmb\phi_a$:
\begin{equation}
\phi_a^{\mu}=\sqrt{12 \rho N\over \lambda}\ \delta^{\mu1}\ .
\label{1replica_config}
\end{equation}
The $1$-replica matrices are then diagonal and one introduces the longitudinal ($\mu =\nu=1$) and transverse components ($\mu =\nu \ne1$). For instance,
\begin{equation}
\begin{array}{ll}
\displaystyle{\widehat{G}_{0L}^{[0]-1}}(q;\rho)&=\displaystyle[{\widehat{G}_0^{[0]-1}}]_{11}(q;\pmb\phi_a)=q^2+m^2+ 6 \rho,\\
\\
\displaystyle{\widehat{G}_{0T}^{[0]-1}}(q;\rho)&=\displaystyle[{\widehat{G}_0^{[0]-1}}]_{\mu\mu;\mu>1}(q;\pmb\phi_a)={q}^2+m^2+ 2\rho.
\end{array}
\label{G0}
\end{equation}
For notational simplicity, we drop the superscript $[0]$ in the rest of this appendix. The Schwinger-Dyson equations for $\widehat{G}_{L}(q;\rho)^{-1}=\widehat{C}_{L}(q;\rho)$ and $\widehat{G}_{T}(q;\rho)^{-1}=\widehat{C}_{T}(q;\rho)$ can now be written as
\begin{equation}
\begin{array}{ll}
&\displaystyle{\widehat{G}_{L}}(q;\rho)^{-1}=q^2+m^2+ 6 \rho+ {\lambda\over  6 N} \int_{q'}  \  \bigg \{\widehat{G}_{L}(q';\rho)  \\
\\
&\displaystyle + \widetilde{G}_{L}(q';\rho) +(N-1)\left(\widehat{G}_{T}(q';\rho) + \widetilde{G}_{T}(q';\rho)\right) + \\
 \\
&\displaystyle {\lambda\over 3N}\int_{q'}\  \widehat{V}(q';\rho) \bigg\{\left(1- \widehat{V}(q';\rho)\,  \widetilde{\Pi}(q';\rho) \right)\widehat{G}_{L}(q+q';\rho)\\
\\
&\displaystyle +  \widetilde{G}_{L}(q+q';\rho)\bigg\} -{4\lambda\over  3N}\ \rho \int_{q'} \ \widehat{V}(q';\rho)^2 \times \\
\\
&\displaystyle \bigg\{\left(1-2\ \widehat{V}(q';\rho)\   \widetilde{\Pi}(q';\rho) \right)\widehat{G}_{L}(q';\rho)\ \widehat{G}_{L}(q+q';\rho) \\
\\
&\displaystyle + \widetilde{G}_{L}(q';\rho) \widehat{G}_{L}(q+q';\rho) +\widehat{G}_{L}(q';\rho) \widetilde{G}_{L}(q+q';\rho)  \bigg\}  \\
\\
&\displaystyle - 4 \rho \  \widehat{V}(q;\rho) \  \widehat{\Pi}(q;\rho)
\label{derivegamma2dagpara}
\end{array}
\end{equation}
and
\begin{equation}
\begin{array}{ll}
&\displaystyle{\widehat{G}_{T}}(q;\rho)^{-1}=q^2+m^2+ 2 \rho +  {\lambda\over  6 N}  \int_{q'} \  \bigg\{\widehat{G}_{L}(q';\rho)\\
\\
&\displaystyle + \widetilde{G}_{L}(q';\rho) +(N-1)\left( \widehat{G}_{T}(q';\rho)+  \widetilde{G}_{T}(q';\rho) \right) +\\
 \\
&\displaystyle {\lambda\over 3N}\int_{q'}\  \widehat{V}(q';\rho) \bigg\{\left(1- \widehat{V}(q';\rho)\,   \widetilde{\Pi}(q';\rho) \right)\widehat{G}_{T}(q+q';\rho)\\
\\
&\displaystyle +  \widetilde{G}_{T}(q+q';\rho)\bigg\} -{4\lambda\over  3N}\ \rho \int _{q'}  \widehat{V}(q';\rho)^2 \times \\
\\
&\displaystyle \bigg\{\left(1-2\  \widehat{V}(q';\rho)\   \widetilde{\Pi}(q';\rho) \right)\widehat{G}_{L}(q';\rho) \widehat{G}_{T}(q+q';\rho) \\
\\
&\displaystyle + \widehat{G}_{L}(q';\rho) \widetilde{G}_{T}(q+q';\rho) +  \widetilde{G}_{L}(q';\rho) \widehat{G}_{T}(q+q';\rho)  \bigg\},
\label{derivegamma2dagortho}
\end{array}
\end{equation}
where the polarization $\widehat{\Pi}(q;\rho)$ is given by 

\begin{equation}
\begin{array}{ll}
&\displaystyle \widehat{\Pi}(q;\rho)={\lambda\over 6N} \int_{q'}\ \bigg\{\widehat{G}_{L}(q';\rho)\widehat{G}_{L}(q+q';\rho)+\\
\\
&\displaystyle (N-1) \widehat{G}_{T}(q';\rho) \widehat{G}_{T}(q+q';\rho) + 2\widehat{G}_{L}(q';\rho) \widetilde{G}_{L}(q+q',\rho)\\
\\
&\displaystyle +2(N-1)  \widehat{G}_{T}(q';\rho)\widetilde{G}_{T}(q+q',\rho)\bigg\}.
\label{eq_hatPi}
\end{array}
\end{equation}
The expressions of $\widetilde{G}_{L}(q;\rho)$ and  $\widetilde{G}_{T}(q;\rho)$, which are 2-replica quantities evaluated for equal replica fields given by Eq.~(\ref{1replica_config}), will be given further below.

We next consider the 2-replica parts and choose a configuration of the two replica fields $\pmb\phi_a$ and $\pmb\phi_b$ parametrized by
\begin{equation}
\begin{array}{ll}
&\displaystyle \phi_a^{\mu}=\sqrt{12 \rho N\over \lambda}\ \delta^{\mu1}\ , \\
\\
&\displaystyle \phi_b^{\mu}=\sqrt{12 \rho' N\over \lambda}\left(\cos\theta\ \delta^{\mu1}+\sin\theta\ \delta^{\mu2}\right).
\label{2replic}
\end{array}
\end{equation}
With this choice, the bare 1-replica matrices $\widehat{\pmb C}_0(q;\pmb \phi_a)$ and $\widehat{\pmb G}_0(q;\pmb \phi_a)$ are still diagonal with longitudinal ($11$) and transverse ($\mu\mu$ with $\mu >1$) components given by Eq.~(\ref{G0}); $\widehat{\pmb C}_0(q;\pmb \phi_b)$ and $\widehat{\pmb G}_0(q;\pmb \phi_b)$ are not diagonal but can be simply diagonalized by applying a transformation  matrix corresponding to a rotation of angle $\theta$,
\begin{equation}
\pmb R_{\theta}=
\left(
\begin{array}{ccc}
 \cos\theta & -\sin\theta  & \pmb 0 \\
\sin\theta & \cos\theta   & \pmb 0 \\
 \pmb 0         &     \pmb 0        &  \pmb\nbOne_{N-2}
\end{array}
\right)\ .
\end{equation}

For instance, 
\begin{equation}
\widehat{\pmb C}_0(q;\pmb \phi_b)= \pmb R_{\theta}\ \widehat{\pmb C}_0^{(diag)}(q;\rho')\  \pmb R_{-\theta},
\end{equation}
with $\widehat{\pmb C}_{0, 11}^{(diag)}(q;\rho')=\widehat{C}_{0L}(q;\rho')$ and $\widehat{\pmb C}_{0, \mu \mu}^{(diag)}(q;\rho')=\widehat{C}_{0T}(q;\rho')$ when $\mu >1$ (and similarly for $\widehat{\pmb G}_0(q;\pmb \phi_b)$). As for $\widetilde{\pmb C}_0(q;\pmb \phi_a, \pmb \phi_b)$, it is simply given by $- \Delta\ \pmb\nbOne$, so that
\begin{equation}
\widetilde{\pmb G}_0(q;\pmb \phi_a, \pmb \phi_b)= - \Delta\ \widehat{\pmb G}_0^{(diag)}(q;\rho)\ \pmb R_{\theta}\ \widehat{\pmb G}_0^{(diag)}(q;\rho')\  \pmb R_{-\theta}.
\end{equation}

The vectorial structure of the  correlation functions is conserved when  fluctuations are taken into account. Thus,  the previous relations are still true when the bare correlation functions are replaced by the ``dressed'' ones. In particular, it is convenient to consider $\widetilde{\pmb G}\ \pmb R_{\theta}$ and $\widetilde{\pmb C}\ \pmb R_{\theta}$ which are related through
\begin{equation}
\begin{split}
\widetilde{\pmb G}&(q;\pmb \phi_a, \pmb \phi_b)\ \pmb R_{\theta}= \\& - \widehat{\pmb G}^{(diag)}(q;\rho)\ \widetilde{\pmb C}(q;\pmb \phi_a, \pmb \phi_b)\ \pmb R_{\theta}\ \widehat{\pmb G}^{(diag)}(q;\rho'),
\end{split}
\end{equation}
and we introduce the components $\widetilde{G}_{LL}=( \widetilde{\pmb G}\ \pmb R_{\theta})_{11}$, $\widetilde{G}_{LA}=( \widetilde{\pmb G}\ \pmb R_{\theta})_{12}$, $\widetilde{G}_{AL}=( \widetilde{\pmb G}\ \pmb R_{\theta})_{21}$, $\widetilde{G}_{AA}=( \widetilde{\pmb G}\ \pmb R_{\theta})_{22}$, $\widetilde{G}_{T}=( \widetilde{\pmb G}\ \pmb R_{\theta})_{\mu \mu;\mu>2}$ (and similarly for $\widetilde{\pmb C}$), so that
\begin{equation}
\begin{array}{ll}
&\widetilde{G}_{LL}(q;\rho,\rho',\theta)=- \widehat{G}_{L}(q;\rho)\,\widehat{G}_{L}(q;\rho')\,\widetilde{C}_{LL}(q;\rho,\rho',\theta)\\
\\
&\widetilde{G}_{LA}(q;\rho,\rho',\theta)=- \widehat{G}_{L}(q;\rho)\,\widehat{G}_{T}(q;\rho')\,\widetilde{C}_{LA}(q;\rho,\rho',\theta)\\
\\
&\widetilde{G}_{AL}(q;\rho,\rho',\theta)=- \widehat{G}_{T}(q;\rho)\,\widehat{G}_{L}(q;\rho')\,\widetilde{C}_{AL}(q;\rho,\rho',\theta)\\
\\
&\widetilde{G}_{AA}(q;\rho,\rho',\theta)=- \widehat{G}_{T}(q;\rho)\,\widehat{G}_{T}(q;\rho')\,\widetilde{C}_{AA}(q;\rho,\rho',\theta)\\
\\
&\widetilde{G}_{T}(q;\rho,\rho',\theta)=- \widehat{G}_{T}(q;\rho)\,\widehat{G}_{T}(q;\rho')\,\widetilde{C}_{T}(q;\rho,\rho',\theta).
\end{array}
\end{equation}
One actually realizes that the description can be further simplified and that only three distinct 2-replica functions are needed,  $\widetilde{G}_{L}(q;\rho,\rho',z) \equiv \widetilde{G}_{LL}(q;\rho,\rho',\theta)$,  $\widetilde{G}_{T}(q;\rho,\rho',z)\equiv \widetilde{G}_{T}(q;\rho,\rho',\theta)$, with $z= \cos \theta$, as well as $\widetilde{G}_{A}(q;\rho,\rho',z)$ which is such that
\begin{equation}
\begin{array}{ll}
&\widetilde{G}_{LA}(q;\rho,\rho',\theta)=\widetilde{G}_{AL}(q;\rho',\rho,-\theta)= \sin \theta\ \widetilde{G}_{A}(q;\rho,\rho',z)\\
\\
&\widetilde{G}_{AA}(q;\rho,\rho',\theta)=\cos \theta\ \widetilde{G}_{A}(q;\rho,\rho',z).
\end{array}
\end{equation}

The Schwinger-Dyson equations can finally be cast in the following compact form:
\begin{equation}
\begin{array}{ll}
&\displaystyle \int_{q'} \bigg \{\widehat{G}_{L}(q;\rho)^{-1}\ \widehat{G}_{L}(q;\rho')^{-1}\ \delta(q+q') -{\lambda\over 3N}\times \\
\\
&\displaystyle Q(q+q';\rho,\rho',z) \bigg \} \widetilde{G}_{L}(q';\rho,\rho',z)- 4 \sqrt{\rho\rho'}\  \widehat{V}(q;\rho)\times \\
\\
&\displaystyle \widehat{V}(q;\rho')\widetilde{\Pi}(q;\rho,\rho',z) = \Delta z,
\label{eqL}
\end{array}
\end{equation}
\\
\begin{equation}
\begin{array}{ll}
&\displaystyle \int_{q'} \bigg \{\widehat{G}_{L}(q;\rho)^{-1}\ \widehat{G}_{T}(q;\rho')^{-1}\ \delta(q+q') -{\lambda\over 3N}\times \\
\\
&\displaystyle Q(q+q';\rho,\rho',z) \bigg \} \widetilde{G}_{A}(q';\rho,\rho',z) = \Delta,
\label{eqA}
\end{array}
\end{equation}
\\
\begin{equation}
\begin{array}{ll}
&\displaystyle \int_{q'} \bigg \{\widehat{G}_{T}(q;\rho)^{-1}\ \widehat{G}_{T}(q;\rho')^{-1}\ \delta(q+q') -{\lambda\over 3N}\times \\
\\
&\displaystyle Q(q+q';\rho,\rho',z) \bigg \} \widetilde{G}_{T}(q';\rho,\rho',z) = \Delta,
\label{eqT}
\end{array}
\end{equation}
with
\begin{equation}
\begin{array}{ll}
&Q(q;\rho,\rho',z)= \widehat{V}(q;\rho)\ \widehat{V}(q;\rho')\bigg\{ \widetilde{\Pi}(q;\rho,\rho',z)\times\\
\\
&\left[1-4 \rho\ \widehat{G}_{L}(q;\rho)\ \widehat{V}(q;\rho) -4 \rho' \ \widehat{G}_{L}(q;\rho')\ \widehat{V}(q;\rho')\right]\\
\\
& + 4\ \sqrt{\rho\rho'}\ \widetilde{G}_{L}(q;\rho,\rho',z)\bigg\}
\label{eq_Q}
\end{array}
\end{equation}
and
\begin{equation}
\begin{array}{ll}
&\displaystyle \widetilde{\Pi}(q;\rho,\rho',z)={\lambda\over 6N} \int_{q'}\  \bigg\{(N-2+z^2)\ \widetilde{G}_{T}(q';\rho,\rho',z)\times\\
\\
&\widetilde{G}_{T}(q+q';\rho,\rho',z)+ \widetilde{G}_{L}(q';\rho,\rho',z)\ \widetilde{G}_{L}(q+q';\rho,\rho',z)+\\
\\
&(1-z) \bigg[\widetilde{G}_{A}(q';\rho,\rho',z)\ \widetilde{G}_{A}(q+q';\rho,\rho',z)+\\
\\
&\widetilde{G}_{A}(q';\rho',\rho,z)\ \widetilde{G}_{A}(q+q';\rho',\rho,z)\bigg]\bigg\}.
\label{eq_tildePi}
\end{array}
\end{equation}
The previously introduced functions of one replica field $\widetilde{G}_{L}(q;\rho)$, $\widetilde{G}_{T}(q;\rho)$, and $\widetilde{\Pi}(q;\rho)$ (see Eqs.~(\ref{derivegamma2dagpara}, \ref{derivegamma2dagortho}, \ref{eq_hatPi})) are simply given by $\widetilde{G}_{L}(q;\rho)\equiv \widetilde{G}_{L}(q;\rho,\rho,1)$, $\widetilde{G}_{T}(q;\rho)\equiv \widetilde{G}_{T}(q;\rho,\rho,1)$, and $\widetilde{\Pi}(q;\rho)\equiv \widetilde{\Pi}(q;\rho,\rho,1)$.

\subsection{Limit of existence of analytic solutions}
Note that in the limit where the two replicas are equal, \textit{i.e.} $\rho=\rho'$ and $z=1$ ($\theta=0$), and provided that all functions are regular enough around this limit, one obtains from Eqs.~(\ref{derivegamma2dagpara}- \ref{eq_hatPi}) and (\ref{eqL}-\ref{eq_tildePi}) a closed set of four coupled self-consistent equations for $\widehat{G}_{L}(q;\rho)$, $\widehat{G}_{T}(q;\rho)$, $\widetilde{G}_{L}(q;\rho)$ and $\widetilde{G}_{T}(q;\rho)$. However, as stressed in the main text, this breaks down if the field dependence is nonanalytic when the two replicas become equal. In the present $O(N)$ model, the limit of equal replica fields can be approached along several directions: one may for instance consider the dependence of the 2-replica functions on $(1-z)$ for $\rho$ and $\rho'$ fixed and on $(\sqrt{\rho}-\sqrt{\rho'})$ for $z$ and $(\sqrt{\rho}+\sqrt{\rho'})$ fixed. (Note that the 2-replica functions are even in the variable $(\sqrt{\rho}-\sqrt{\rho'})$ but not in $(1-z)$.) To generalize what has been done in section IV-B for a one-component field, we apply $\partial_{y}=\partial/\partial y$ on eqs.~(\ref{eqL}-\ref{eqT}) with $y=z$ or $y=\sqrt{\rho}-\sqrt{\rho'}$. So, for instance, one finds
\begin{equation}
\begin{array}{ll}
&\displaystyle \partial_{y}Q(q;\rho,\rho',z)\big |_{\pmb \phi}= 4 \rho\ \widehat{V}(q;\rho)^2\  \partial_{y}\widetilde{G}_{L}(q;\rho,\rho',z)\big |_{\pmb \phi} +\\
\\
&\displaystyle  \frac{\lambda}{3}\  \widehat{V}(q;\rho)^2\left[1- 8 \rho\ \widehat{G}_{L}(q;\rho)\ \widehat{V}(q;\rho)\right] \int_{q'} \widetilde{G}_T(q+q';\rho)\times \\
\\
& \partial_{y}\widetilde{G}_{T}(q';\rho,\rho',z)\big |_{\pmb \phi} + \displaystyle O({1\over N}) + \rm{reg},
\end{array}
\end{equation}
where the subscript $\pmb \phi$ indicates that the derivatives are evaluated for equal replica fields, $\rho=\rho'$ and $z=1$, and ``reg'' denotes regular terms that involve derivatives of 1-replica quantities.

Due to the presence of three different 2-replica functions, $\widetilde{G}_{L}$, $\widetilde{G}_{A}$, $\widetilde{G}_{T}$, the operator $\Lambda^{[0]}(q,q')$ is now a matrix with several distinct components. To avoid too much formalism, we derive here the operator in the case where $\rho=\rho'=0$ (and of course $z=1$). Generalization is tedious but straightforward. To find the potential breakdown of analyticity, we evaluate the operator under the assumption that the 2-replica functions $\widetilde{G}_{L}$, $\widetilde{G}_{A}$, $\widetilde{G}_{T}$, obtained as solutions of the above Schwinger-Dyson equations, are analytic in the limit of equal replica fields. It is then easy to realize that $\widehat{G}_{L}=\widehat{G}_{T}\equiv \widehat{G}$, $\widetilde{G}_{L}=\widetilde{G}_{A}=\widetilde{G}_{T}\equiv \widetilde{G}$ and that, as a result, the operator $\Lambda^{[0]}(q,q')$ has a single component corresponding to the equation
\begin{equation}
\int_{q'} \Lambda^{[0]}(q,q') \partial_{y}\widetilde{G}(q';\rho,\rho',z)\big |_{\pmb \phi=\pmb0} = \rm{reg},
\end{equation}
with ``reg'' equal to zero if $y=\sqrt{\rho}-\sqrt{\rho'}$ (due to the permutation symmetry between replicas) and ``reg'' finite if $y=z$. The operator $\Lambda^{[0]}(q,q')$ is found equal to
\begin{equation}
\begin{array}{ll}
&\Lambda^{[0]}(q,q')=\displaystyle  \widehat G(q;\rho=0)^{-2}\ \delta(q+q')-\\
\\&\displaystyle {\lambda \over 3N} Q(q+q';0,0,1) 
- \left( {\lambda \over 3}\right) ^2{1 \over N}\int_{q''}\ \widehat V(q'';\rho=0) ^2 \\
\\&\displaystyle \times  \widetilde G(q+q'';\rho=0)\ \widetilde G(q''+q';\rho=0)+ O(\frac{1}{N^2}).
\end{array}
\label{replicon1/N}
\end{equation}
In addition, from Eqs.~(\ref{eq_Q}) and (\ref{eq_tildePi}), one has
\begin{equation}
\begin{array}{ll}
&\displaystyle Q(q;0,0,1)= \widehat V(q;\rho=0) ^2\ \widetilde{\Pi}(q;\rho=0),
\end{array}
\end{equation}
\begin{equation}
\begin{array}{ll}
&\displaystyle \widetilde{\Pi}(q;\rho=0)={\lambda \over 6}\int_{q'} \widetilde G(q';\rho=0)\ \widetilde G(q+q';\rho=0),
\end{array}
\end{equation}
with $\widehat V(q;\rho=0)=(1+\widehat{\Pi}(q;\rho=0))^{-1}$, and
\begin{equation}
\begin{array}{ll}
\displaystyle \widehat{\Pi}(q;\rho=0)&=\displaystyle {\lambda \over 6}\int_{q'}\ \widehat G(q';\rho=0)\times \\
\\&\displaystyle \left(\widehat G(q+q';\rho=0)+2\ \widetilde G(q+q';\rho=0) \right).
\end{array}
\end{equation}
\\
It is now easy to check that the operator $\Lambda^{[0]}(q,q')$, which is associated with the limit of validity of the analytic solutions to the Schwinger-Dyson equations, coincides with the replicon operator derived in Refs.~[\onlinecite{mezard94,dedominicis95}].

\section{Stability operator for local self-energies}

In several approximations or models (mean field, Gaussian variational method, locator and related DMFT approximations), the self-energies are taken as purely local quantitites. As a result of the Schwinger-Dyson equation, Eq.~(\ref{GG}), the only momentum dependence in the (dressed) propagators comes from their bare counterparts. For instance, in the explicit RSB scheme (considering for simplicity a 1-component field), one has
\begin{equation}
\widehat{G}^{[0]}(q;\phi)=\left[ \widehat{G}_0^{[0]-1}(q;\phi)- \widehat{\Sigma}^{[0]}(\phi)\right] ^{-1},
\label{widehatG0local}
\end{equation}
\begin{equation}
\begin{split}
&\widetilde{G}^{[0]}(q;\phi,\phi') = \\&
\\ 
&\displaystyle - \widehat{G}^{[0]}(q;\phi) \left[ \widetilde{G}_0^{[0]-1}(q;\phi,\phi') -  \widetilde{\Sigma}^{[0]}(\phi,\phi')\right] \widehat{G}^{[0]}(q;\phi').
\end{split}
\label{widetildeG0local}
\end{equation}
By assumption, the self-energies are functions of the local piece of the propagators only, \textit{e.g.} $\widetilde{\Sigma}^{[0]}(\phi,\phi')$ only depends on $\widetilde{G}^{[0]}(x=0;\phi,\phi')=\int_q \widetilde{G}^{[0]}(q;\phi,\phi')$. 

From Eq.~(\ref{lambda0}), the stability operator $\Lambda^{[0]}(q,q';\phi,\phi)=\Lambda_R(q,q')$ is then given by
\begin{equation}
\Lambda_R(q,q')=\widehat{G}(q;\phi)^{-2}\delta(q+q') - \frac{\delta \widetilde{\Sigma}(\phi,\phi')}{\delta \widetilde{G}(x=0;\phi,\phi')}\bigg|_{\phi},
\label{replicon_local}
\end{equation}
where the last term has no momentum dependence and is evaluated for $\phi'=\phi$.

Finding the eigenvalues of this stability operator is now an easy task. There is in fact a single eigenvalue $\lambda_R$ which is solution of
\begin{equation}
\displaystyle 1=\frac{\delta \widetilde{\Sigma}}{\delta \widetilde{G}}\bigg|_{\phi} \int_q\ \frac{\widehat{G}(q;\phi)^2}{1-\lambda_R \widehat{G}(q;\phi)^2},
\end{equation}
with the associated eigenfunction $\Psi_R(q;\phi)\propto \widehat{G}(q;\phi)^2 (1-\lambda_R \widehat{G}(q;\phi)^2)^{-1}$ (above and in what follows, we drop the superscript $[0]$). The operator becomes marginal when $\lambda_R=0$, which characterizes a generalized Almeida-Thouless line \cite{almeida78} and happens whenever $1=\delta \widetilde{\Sigma}/ \delta \widetilde{G}|_{\phi} \int_q \widehat{G}(q;\phi)^2$.

The inverse operator $\chi_R(q,q')$ is obtained from Eq.~(\ref{replicon_local}) as
\begin{equation}
\displaystyle \chi_R(q,q') =\widehat{G}(q)^2 \left[ \delta(q+q') +  \frac{\frac{\delta \widetilde{\Sigma}}{\delta \widetilde{G}}\big|_{\phi}}{1-\frac{\delta \widetilde{\Sigma}}{\delta \widetilde{G}}\big|_{\phi} \int_q \widehat{G}(q)^2} \widehat{G}(q')^2 \right]
\end{equation}
and the spin-glass susceptibility, $\chi_{SG} = \int_q \int_{q'}\chi_R(q,q')$, is simply given by
\begin{equation}
\displaystyle \chi_{SG}=  \displaystyle {\int_q \widehat{G}(q)^2 \over 1- \frac{\delta \widetilde{\Sigma}}{\delta \widetilde{G}}\big|_{\phi} \int_q \widehat{G}(q)^2}.
\end{equation}
In this case, which encompasses most approximations through which spontaneous RSB has been found, the spin-glass susceptibility diverges as soon as the replicon eigenvalue $\lambda_R$ goes to zero.

\end{document}